\documentclass[11pt]{article}
\usepackage[centertags]{amsmath}
\usepackage{amsfonts}
\usepackage{amssymb}
\usepackage{color}
\usepackage{graphicx}
\usepackage{cite}
\numberwithin{equation}{section}
\numberwithin{figure}{section}
\numberwithin{table}{section}

\newcommand{\abs}[1]{| #1 |}
\newcommand{\Abs}[1]{\left| #1 \right|}

\newcommand{\tr}{\mathrm{tr}}
\newcommand{\sgn}{\mathrm{sgn}}

\newcommand{\bu}{{\bar u}}

\newcommand{\cA}{{\cal A}}

\newcommand{\cE}{{\cal E}}

\newcommand{\cN}{{\cal N}}
\newcommand{\cS}{{\cal S}}

\newcommand{\Z}{\mathbb{Z}}

\newcommand{\R}{\mathbb{R}}

\newcommand{\act}[1]{\mathrm{S}_{#1}}
\newcommand{\dd}{\mathrm{d}}

\newcommand{\du}{\partial_u}

\newcommand{\dv}{\partial_v}

\newcommand{\dw}{\partial_w}

\newcommand{\dz}{\partial_z}

\newcommand{\ap}{{\alpha'}}

\newcommand{\sqda}{\sqrt{2 \alpha'}}
\newcommand{\sqha}{\sqrt{ \frac {\alpha'} {2} } }
\newcommand{\xzm}{{x_0}}
\newcommand{\oh}{\frac{1}{2}}

\newcommand{\BO}{BO }
\newcommand{\NBO}{NBO }
\newcommand{\GNBO}{GNBO }

\newcommand{\A}{ A}
\newcommand{\BB}{ B(u) }

\newcommand{\Ki}[1]{{\cal K}^{#1}}

\newcommand{\xp}{ {x^+} }
\newcommand{\xm}{ {x^-} }
\newcommand{\xx}{ {x^2} }
\newcommand{\xxx}{ {x^3} }
\newcommand{\vex}{ {\vec x} }
\newcommand{\hD}{ (2\pi \Delta) }

\newcommand{\epp}{ {\epsilon_+} }
\newcommand{\epm}{ {\epsilon_-} }
\newcommand{\epd}{ {\epsilon_2} }
\newcommand{\epu}{ {\epsilon_u} }
\newcommand{\epv}{ {\epsilon_v} }
\newcommand{\epz}{ {\epsilon_z} }

\newcommand{\eee}{ {\epp\, \epm\, \epd } }
\newcommand{\eppN}[1]{ {\epsilon_{(#1) +}} }
\newcommand{\epmN}[1]{ {\epsilon_{(#1) -}} }
\newcommand{\epdN}[1]{ {\epsilon_{(#1) 2}} }
\newcommand{\epN}[1]{ {\epsilon_{(#1)}} }

\newcommand{\Ss}[2]{ { S_{#1\, #2} } }
\newcommand{\SsN}[1]{ { S_{(#1)} } }

\newcommand{\mm}{ {l} }
\newcommand{\kp}{ {k_+} }
\newcommand{\km}{ {k_-} }
\newcommand{\kk}{ {k_2} }
\newcommand{\kw}{ {p} }
\newcommand{\vk}{ {\vec k} }
\newcommand{\mmN}[1]{ {l_{(#1)}} }

\newcommand{\mN}[1]{ {m_{(#1)}} }

\newcommand{\kwN}[1]{ {p_{(#1)}} }
\newcommand{\kpN}[1]{{ {k_{(#1) +}} }}
\newcommand{\kmN}[1]{{ {k_{(#1) -}} }}
\newcommand{\kkN}[1]{{ {k_{(#1) 2}} }}
\newcommand{\vkN}[1]{{ {\vec k_{(#1)}} }}
\newcommand{\rN}[1]{ {r_{(#1)}} }
\newcommand{\kN}[1]{ {k_{(#1)}} }

\newcommand{\kkk}{ {\kp\, \km\, \kk } }
\newcommand{\kkkN}[1]{ {\kpN{#1}\, \kmN{#1}\, \kkN{#1} } }
\newcommand{\kkkk}{ {\kp\, \km\, \kk\, \vk } }

\newcommand{\kmkr}{ { \{\kp\, \mm\, \vec k\, r  \}  } }
\newcommand{\MINUSkmkr}{ { \{-\kp\, -\mm\, -\vec k\, r  \}  } }
\newcommand{\kmkrgen}{ { \{\kp\, \kw\, \mm\, \vec k\, r  \}  } }

\newcommand{\kmkrN}[1]{ { \{ \kpN{#1}\, \mmN{#1}\, \vkN{#1}\, \rN{#1} \} }}
\newcommand{\kmkrgenN}[1]{ { \{ \kpN{#1}\, \kwN{#1}\, \mmN{#1}\, \vkN{#1}\, \rN{#1} \} }}
\newcommand{\MINUSkmkrN}[1]{ { \{ -\kpN{#1}\, -\mmN{#1}\, -\vkN{#1}\, \rN{#1} \} }}
\newcommand{\MINUSkmkrgenN}[1]{ { \{ -\kpN{#1}\, -kwN{#1}\, -\mmN{#1}\, -\vkN{#1}\, \rN{#1} \} }}

\newcommand{\kmrN}[1]{ { \{ \kpN{#1}\, \kmN{#1}\, \mmN{#1}\,  \rN{#1} \} }}

\newcommand{\tpkmkr}{ {\tilde \phi}_\kmkr }
\newcommand{\tpkmkrgen}{ {\tilde \phi}_\kmkrgen }

\newcommand{\tpkmkrN}[1]{ {\tilde \phi}_{\kmkrN{#1}} }
\newcommand{\tpkmkrgenN}[1]{ {\tilde \phi}_{\kmkrgenN{#1}} }
\newcommand{\pkmrN}[1]{ {\phi}_{\kmrN{#1}} }
\newcommand{\tponlyN}[1]{ {\tilde \phi}_{ (#1) } }
\newcommand{\pkmkr}{ {\phi}_\kmkr }
\newcommand{\pkmkrN}[1]{ {\phi}_{\kmkrN{#1}} }

\newcommand{\takmkrgen}[1]{ {\tilde a}_{\kmkrgen \, #1 } }

\newcommand{\cNE}[1]{{\cal E}_{ \kmkr \, \underline{#1} } }
\newcommand{\cEN}[1]{ {\cal E} _ {\kmkrN{#1} }}

\newcommand{\cNEN}[2]{{\cal E}_{{ \kmkrN{#2} \, \underline{#1} } }}
\newcommand{\cNEgen}[1]{{\cal E}_{ \kmkrgen \, \underline{#1} } }

\newcommand{\INT}[2]{{\cal I}_{\{#1\}} ^{[#2]} }
\newcommand{\MINT}[2]{{\cal J}_{(#1)} ^{[#2]} }
\newcommand{\INText}[2]{{\cal I}_{#1} ^{[#2]} }

\newcommand{\Dx}{\Delta_2}
\newcommand{\Dxx}{\Delta_3}

\newcommand{\cNBO}{ {\cN}_{BO} }
\newcommand{\vvphi}{ \varphi}
\newcommand{\LLP}[1]{ {e^{#1  \Delta {\vvphi} } } }
\newcommand{\LLB}[1]{ { e^{ #1  \Delta {\beta} } } }
\newcommand{\LLPB}[1]{ { e^{#1  \Delta ({\vvphi+\beta}) } } }

\newcommand{\loLd}{ \Delta }

\newcommand{\nsm}{ {\sigma_-} }

\newcommand{\vsm}{ {\varsigma_-} }
\newcommand{\hspm}{ {\hat \sigma_{-}} }

\newcommand{\BOkk}{ \kp \km }
\newcommand{\BOkkN}[1]{ \kp_{(#1)} \km_{(#1)} }

\newcommand{\lsi}{ l\, \hspm }
\newcommand{\lsiN}[1]{ l_{(#1)}\, \hspm{}_{(#1)} }
\newcommand{\tplsi}{ {\tilde \phi}_{\lsi} }
\newcommand{\plsiN}[1]{ {\phi}_{ \lsiN{#1} } }

\newcommand{\lkrsi}{ {l\, \vk\, r\, \nsm } }
\newcommand{\HH}{ H_\lkrsi }
\newcommand{\ml}{m l\,}

\newcommand{\KKK}[2]{ k_{ (#1)\, #2 } }     

\newcommand{\COMMENTO}[1]{}

\begin{document}

\title{
On the Origin of Divergences in Time-Dependent Orbifolds
}

\author{
Andrea Arduino,\, Riccardo Finotello,\, Igor Pesando
\\ \\
Dipartimento di Fisica, Universit\`{a} di Torino \\
and I.N.F.N. - sezione di Torino \\
Via P. Giuria 1, I-10125 Torino, Italy
\\
\vspace{0.3cm}
\\
andrea.arduino@to.infn.it \\
riccardo.finotello@to.infn.it \\
ipesando@to.infn.it}

\maketitle
\thispagestyle{empty}

\abstract{
We consider time-dependent orbifolds in String Theory and 
we show that divergences are not associated with a gravitational
backreaction since they appear in the open string sector too.
They are related to the non existence of the underlying effective field theory as in several cases fourth and higher order contact terms do not exist.

Since contact terms may arise from the exchange of string massive states,
we investigate and show that some three points
amplitudes with one massive state in the open string sector are divergent on the time-dependent orbifolds.

To check that divergences are associated with the existence of a discrete zero eigenvalue of the Laplacian of the subspace with vanishing volume,
we construct the Generalized Null Boost Orbifold where this phenomenon can be turned on and off.

}

keywords: {String theory, QFT in curved space}

\newpage

\tableofcontents

\newpage


%
%

\section{Introduction and Conclusion}

  String Theory is often considered to be one, if not the best, candidate to
  describe quantum gravity and therefore the Big Bang singularity.
  Unfortunately and puzzlingly, the first attempts to consider
  space-like \cite{Craps:2002ii} or light-like singularities \cite{Liu:2002ft}
  by means of orbifold  techniques yielded divergent four points
  {\sl closed string} amplitudes (see
  \cite{Cornalba:2003kd,Craps:2006yb} for reviews).
  This is somewhat embarrassing for a theory touted as a theory of
  quantum gravity. The aim of this paper is to elucidate the origin of these
singularities in amplitudes.

A number of reasons are cited in the literature for the existence of
these singularities.
The most widespread is that they are due to large backreaction of the
incoming matter into the singularity due to the exchange of a single graviton.
However this is false.

What has gone unnoticed is that in the Null Boost Orbifold (NBO)
\cite{Liu:2002ft} even the four
{\sl open string} tachyons amplitude is divergent.
Since we are working at tree level this means that gravity is not the
problem.
In fact in eq.\ (6.16) of \cite{Liu:2002ft}, the four tachyons amplitude in the divergent region reads
\begin{align*}
    A_4 \sim \int_{q\sim 0} \frac{\dd q}{|q|} A(q)
    \quad \mbox{ with } \quad &
    A_{closed}\sim q^{\ap \left( \frac{4}{\ap}- \vec p_{\perp t}^2 \right)}
\\
\quad \mbox{ and } \quad &
A_{open}\sim q^{\ap ( \frac{1}{\ap} - \vec p_{\perp t}^2 ) } 
\, \tr\left(\{T_1,T_2\} \{T_3, T_4\}\right),
\end{align*}
where $T$ are the Chan-Paton matrices.
Moreover divergences in string amplitudes are not limited to four points: interestingly 
we show that the open string three point amplitude with two tachyons and the first
massive state may be divergent when some {\sl physical} polarization are chosen.

The true problem is therefore not related to a gravitational issue, 
but to the non existence of the effective field theory.
In fact when we express the theory using the eigenmodes of the kinetic
terms some coefficients do not exist, even as a distribution.
This is true for both open and closed string sectors since it manifests also in the four scalar contact term.
This problem can be roughly traced back to the vanishing volume of a subspace
and the existence of a discrete zero mode of the Laplacian on this subspace.

In Section~\ref{sect:NOscalarQED} in order
to elucidate the problem of singularities in open string we start by considering the \NBO where
we try to construct a scalar QED theory. 
However, even the four scalars vertex is ill defined.

Divergences in scalar QED are due to the singularity of the
eigenfunctions of the scalar d'Alembertian near the singularity but in a
somehow unexpected way. Near the singularity $u=0$ all but one eigenfunctions
behave as $\frac{1}{\sqrt{\abs{u}}} e^{i \frac{\cA}{u}}$ with $\cA\ne 0$.
The product of $N$ eigenfunctions gives a singularity
$\abs{u}^{-N/2}$ which is technically not integrable. However the exponential term
$e^{i \frac{\cA}{u}}$ allows for an interpretation as distribution when $\cA=0$ is not an isolated point.
When $\cA=0$ is isolated the singularity is definitely not
integrable and there is no obvious interpretation as a distribution.
Specifically in the \NBO $\cA\sim\frac{l^2}{\kp}$ with $l$ the momentum along the compact direction
therefore there is one eigenfunction with isolated $\cA=0$:
the one which is associated with the discrete
momentum $\mm=0$ along the orbifold compact direction.
It is the eigenfunction which is constant along that direction and 
it is the root of all divergences.
If the direction were not compact or there were at least another non compact direction contributing to $\cA$ we could avoid the problem.
We then check whether the most obvious ways of regularizing the theory by making  $\cA$ not vanishing  may work.
The first regularization we try is to use a Wilson line along the compact direction.
It works for the scalar QED and almost for string theory but not completely.
The diverging three point string amplitude involves an anti-commutator of the Chan-Paton factor therefore  it is divergent also for a neutral string, i.e. for a string with both ends attached to the same brane. This kind of string does not feel Wilson lines.
Moreover anti-commutators are present in amplitudes with massive states in unoriented and supersymmetric strings and therefore neither worldsheet parity nor supersymmetry  help.
The second obvious regularization is to use higher derivatives couplings to Ricci tensor which is the only non vanishing tensor associated with the (regularized) metric. Unfortunately if we assume that the parameter entering the metric regularization is of the order of the string scale these terms do not help.
In any case it seems that a sensible regularization must coupe to all open string in the same way and this suggests a gravitational coupling.
It would be interesting to check whether closed string winding modes
could help \cite{Pioline:2003bs}.

In any case we now understand the origin of the string divergences
from the point of view of non existence of contact terms in the effective field theory.
String theory divergences come from ill defined contact
terms which must be reproduced by string amplitudes. 
In the effective field theory
contact terms arise from String Theory also through the exchange of massive string
states and this suggests to examine three point amplitudes with one massive state.

To do so, even if not strictly necessary to show the existence of divergences, we want to understand how to write the polarizations for the massive state on orbifold  from the Minkowski ones. 
This is tackled in Section~\ref{sect:Eigenmodes_from_Covering}.

In Section~\ref{sec:overlap} we consider overlaps of different wave functions and derivatives thereof.
These overlaps are related in fact to the coefficients of the expansions of the effective theory in eigenfunctions of the kinetic terms and therefore they are strictly related to string amplitudes on orbifold.

In Section~\ref{sect:NO3ptsMassive}
we go back to String Theory and we use the result of the previous section
in order to verify that in the \NBO the
open string three points amplitude with two tachyons and one first level
massive string state does indeed diverge when some physical polarization are
chosen.
An intuitive reason is that we have an infinite number of images and
the delta functions associated with momentum conservation have an
accumulation point of their support.
Nevertheless the existence of the accumulation point is not sufficient
since three tachyons amplitude converges: the coefficients of the
deltas matter, too, and the convergence must be verified.

As stated above, if the directions with vanishing volume
were not compact or a ``mixture'' of compact ones and at
least one non compact we could avoid the divergences.
To check this point, in Section~\ref{sect:genNOscalarQED}
we introduce the Generalized Null Boost Orbifold (GNBO) as a generalization of the \NBO which still has a light-like singularity and is generated by one Killing vector.
However in this model there are two directions associated with $\cA$, one
compact and one non compact.
We can then construct the scalar QED and  the effective field theory which extends it
with the inclusion of higher order terms since all terms have a
distributional interpretation.
However if a second Killing vector is used to compactify the formerly non compact direction the theory has again the same problems as in the \NBO case.

In Section~\ref{sect:BO}
we then quickly examine the Boost Orbifold (BO) where the divergences are generically milder.
The scalar eigenfunctions generically behave as $|t|^{\pm i\, \frac{l}{\Delta} }$ near the singularity but there is one eigenfunction which behaves as $\log(|t|)$ and again it is the constant eigenfunction along the compact direction which is the origin of all divergences.
In particular the scalar QED can be defined and
the first term which gives a divergent contribution is of the form 
$\abs{\phi~\dot \phi}^2$, i.e
divergences are hidden into the derivative expansion of the effective field theory.
Again three points open string amplitudes with one massive state diverge.

  
The lessons to be learnt from these examples are many.

First, it seems that String Theory cannot do better than field theory when
the latter does not exist, at least at the perturbative level where one deals with particles.

Second, when spacetime becomes singular, the string massive modes are not anymore spectators.

Third and related to the last point,
the previous problems seem to suggest that issues with spacetime
singularities are hidden into contact terms and interactions with
massive states. This would explain in an intuitive way why the eikonal
approach to gravitational scattering seems to work well: eikonal is in fact
concerned with three point massless interactions.
In fact it appears \cite{NOI2020xx} that the classical and quantum scattering 
on an em wave \cite{Jackiw:1991ck}
or
gravitational wave \cite{tHooft:1987vrq}
in \BO and \NBO are well behaved.
From this point of view the ACV approach \cite{Soldate:1986mk,Amati:1987wq}, especially when considering
massive external states \cite{Black:2011ep}, may be more sensible.

Finally it seems that all issues are related with what happens to the Laplacian associated with the spacelike subspace with vanishing volume at the singularity. If there is a discrete zero eigenvalue the theory develops divergences. 

\section{Scalar QED on \NBO and Divergences} 
\label{sect:NOscalarQED}

As we discussed in the Introduction, the four open string tachyons
amplitude diverges in the \NBO\!\!.
Given the suggestion in the literature \cite{Cornalba:2003kd}
that this can be cured by the
eikonal resummation, we would like to consider the scalar QED on the
\NBO\!\!.
Another reason is that all eigenmodes can be written using elementary
functions thus making the issues more transparent.
Its action is given by
\begin{equation}
  \act{sQED}
  =
  \int_\Omega \dd^D x\,
\sqrt{- \det g}
      \left(
      - (D^\mu \phi)^* D_\mu \phi
      -M^2  \phi^* \phi
    -
    \frac{1}{4} f^{\mu\nu}\,f_{\mu\nu}
    -
    \frac{g_4}{4}
    \abs{\phi}^4
    \right)
,
\end{equation}
with
\begin{equation}
    D_\mu \phi= (\partial_\mu -i\,e\, a_\mu) \phi,
    \qquad
    f_{\mu\nu}=\partial_\mu a_\nu-\partial_\nu a_\mu.
\end{equation}
We reserve small letters for quantities defined on the orbifold
and capital ones for those defined in Minkowski.
Moreover $\Omega$ denotes the orbifold.
We will construct directly both the scalar and the spin-1 eigenfunctions 
which we can use as a starting point for the perturbative computations.

\subsection{Geometric Preliminaries}
\label{sec:geometric_preliminaries_nbo}

In $\R^{1,D-1}$ with coordinates $(x^\mu)=(\xp, \xm, \xx, \vex)$
and metric
\begin{equation}
\dd{s}^2 = - 2 \dd{\xp} \dd{\xm} + (\dd{\xx})^2 + \eta_{ij}\dd{x}^i\dd{x}^j
,
\end{equation}
we consider the
following change of coordinates to $(x^\alpha)=(u, v, z, \vex)$
\begin{align}
\begin{cases}
  \xm &= u
  \\
  \xx&= \Delta u z
  \\
  \xp &= v+ \oh \Delta^2 u z^2
\end{cases}
\Leftrightarrow \,\,\,
\begin{cases}
  u &= \xm
  \\
  z&= \frac{\xx}{\Delta\, \xm}
  \\
  v &= \xp- \oh \frac{(\xx)^2}{\xm}
\end{cases}
.
\label{eq:NBO_coordinates}
\end{align}
Then the metric becomes:
\begin{equation}
  \dd s^2
=
  -2\, \dd u\, \dd v + (\Delta u )^2 (\dd z )^2 + \eta_{ij}\dd{x}^i\dd{x}^j
  ,
\end{equation}
along with the non vanishing geometrical quantities
\begin{equation}
    -\det{g} = ( \Delta u )^2
    ,
\end{equation}
and
\begin{equation}
    \Gamma^v_{z z} = \Delta^2 u,
    \qquad
    \Gamma^z_{u z} = \frac{1}{u}.
\end{equation}
Riemann and Ricci tensor components however vanish since at this stage we only performed a change of coordinates from the original Minkowski spacetime: locally it is the same as the \NBO and they must have the same local differential geometry.

The \NBO is introduced by identifying points along the orbits of the Killing vector:
\begin{equation}
    \begin{split}
        \kappa & = - 2 \pi i \Delta J_{+2}
        \\
               & = 2 \pi \Delta 
               (x^2 \partial_+ + \xm \partial_2)
        \\
               & = 2 \pi \dz
               ,
    \end{split}
\end{equation}
in such a way that
\begin{equation}
    x^{\mu} \equiv  \Ki{n} x^{\mu}, \qquad n \in \Z
,
\end{equation}
where $\Ki{n}= e^{n\kappa}$,
leads to the identifications
\begin{equation}
x=
  \begin{pmatrix}
  \xm \\  \xx \\ \xp \\ \vex
  \end{pmatrix}
  \equiv
  \Ki{n} x
=
  \begin{pmatrix}
   \xm\\
   \xx + n \hD \xm\\
   \xp+ n \hD \xx + \oh n^2 \hD^2 \xm,\\
   \vex
   \end{pmatrix}
 \end{equation}
or to 
\begin{equation}
    ( u, v, z, \vex ) \equiv ( u, v, z + 2 \pi n, \vex )
\end{equation}
in the coordinates $(x^\alpha)$ where $\kappa=2 \pi \partial_z$ is a global Killing vector.

For future use in Section~\ref{sec:saving}, we notice that we could regularize the metric as
 \begin{equation}
  \dd s^2
  =
  -2\, \dd u\, \dd v + \Delta^2 (u^2+\epsilon^2) (\dd z )^2 + \eta_{ij}\dd{x}^i\dd{x}^j
  .
\end{equation}
Then the non vanishing geometrical quantities are
\begin{equation}
  -\det g = \Delta^2 (u^2+\epsilon^2),
\end{equation}
and
\begin{equation}
    \Gamma^v_{z z}=\Delta^2 u,
    \qquad
    \Gamma^z_{u z}= \frac{u}{u^2+ \epsilon^2},
\end{equation}
which lead to the following Riemann and Ricci tensor components:
\begin{equation}
  R^{z}_{~u z u}= - \frac{\epsilon^2}{ (u^2+ \epsilon^2)^2 },
  \quad
  R^{v}_{~z z u}= - \frac{\Delta^2 \epsilon^2}{ u^2+ \epsilon^2 },
  \quad
  Ric_{ u u} = -\frac{\epsilon^2}{ (u^2+ \epsilon^2)^2 }
  .
\end{equation}
  Since
  $\delta_{reg}(u)= \frac{1}{\pi} \frac{\epsilon}{ u^2+ \epsilon^2}$
this means that $R^{z}_{~u z u} \sim [\delta_{reg}(u)]^2$.

 \subsection{Free Scalar Action} 
 
We now want to find the eigenmodes of the Laplacian in order to write in a
diagonal way the scalar kinetic term given by\footnote{
    The factor $-g^{\alpha\beta}$ is due to the choice of the East coast convention for the metric, i.e.:
    \begin{equation*}
    -g^{\alpha\beta} \partial_{\alpha} \phi^* \partial_{\beta} \phi - M^2 \phi^* \phi \sim + \abs{\dot{\phi}}^2 - M^2 \abs{\phi}^2 \sim \mathrm{E}^2 - M^2.
    \end{equation*}
}
\begin{equation}
    \begin{split}
        \act{\text{scalar kin}}
        & =
        \int_\Omega \dd^D x\,
        \sqrt{- \det g}~
          \Bigl(
          -g^{\alpha\beta} \partial_\alpha \phi^* \partial_\beta \phi
          -M^2  \phi^* \phi
          \Bigr)
        \\
        & =
          \int \dd^{D-3} \vec x\,
          \int \dd u\, \int \dd v\, \int_0^{2\pi} \dd z\,
          \abs{ \Delta u}
          \Biggl(
          \partial_u \phi^*\,\partial_v \phi\,
        \\
        & 
        +
          \partial_v \phi^*\,\partial_u \phi\,
          -
          \frac{1}{(\Delta u )^2}  \partial_z \phi^*\,\partial_z \phi\,
          -
          \partial_i \phi^* \partial_i \phi
          -
          M^2 \phi^* \phi
          \Biggr)
    .
    \end{split}
\end{equation}
    The solution to the equation of motion is enough when we want to perform
    the canonical quantization.
    Since we want to use the Feynman diagrams, we consider the
    path integral approach: we take off-shell modes and
    solve the eigenvalue problem $\square \phi_r = r \phi_r$.
    By comparing with the flat case we see that
    $r$ is $2 k_-\kp-\vec k^2$ when $k$ is the flat coordinates
    momentum.
    We therefore have
    \begin{align}
      -2 \partial_u \partial_v \phi_r
      -
      \frac{1}{u} \partial_v \phi_r
      +
      \frac{1}{( \Delta u)^2} \partial_z^2 \phi_r
      +
      \partial_i^2 \phi_r
      =
      r \phi_r
      .
    \end{align}
    Using Fourier transforms, it then easily follows 
    that the eigenmodes are
\begin{equation}
        \phi_\kmkr(u,v,z,\vec x)
        =
        e^{i \kp v +i \mm z+ i \vec k \cdot \vec x}\,
            \tpkmkr(u)
,
\end{equation}
with
      \begin{align}
        \tpkmkr(u)
         &=
        \frac{1}{\sqrt{ ( 2 \pi )^D~ 2|\Delta \kp|~ |u|} }
        e^{
          -i \frac{\mm^2}{2 \Delta^2 \kp} \frac{1}{u}
          +i \frac{\vec k^2+r}{2 \kp} u
        }
        ,
      \end{align}
and
    \begin{equation}
        \phi^*_\kmkr(u,v,z,\vex) =
        \phi_{\MINUSkmkr}(u, v, z, \vex),
    \end{equation}
      where we have chosen the numeric factor in order to get a canonical normalization:
\begin{equation}
    \begin{split}
    &
    ( \phi_{\kmkrN 1},\,  \phi_{\kmkrN 2})
    \\
    & =
    \int \dd^{D-3} \vec x\,
    \int \dd u\, \int \dd v\, \int_0^{2\pi} \dd z~
    \abs{ \Delta u }
     \phi_{\kmkrN 1}\,  \phi_{\kmkrN 2}
  \\
  &
    =
    \delta^{D-3}( \vkN{1} + \vkN{2})\,
    \delta( r{}_{(1)} - r{}_{(2)})\,
    \delta( \kpN{1} + \kpN{2})\,
    \delta_{\mmN{1} , -\mmN{2}}
    .
    \end{split}
\end{equation}
We can then perform the off-shell expansion
    \begin{align}
      \phi(u,v,z,\vec x)
      &=\!
      \int \dd^{D-3} \vec k
      \int \dd r\!
      \int \dd \kp
      \sum_{\mm\in\Z}
        \cA_\kmkr\,
        \phi_\kmkr(u,v,z,\vec x)
            ,
\end{align}
so that the scalar kinetic term becomes 
\begin{align}
    \act{\text{scalar kin}}
    =&
      \int \dd^{D-3} \vec k\,
      \int \dd r
      \int \dd \kp\, 
      \sum_{\mm\in\Z}\,
       (r- M^2)\,
       \cA_\kmkr\,
       \cA_\kmkr^*
       .
  \end{align}

  \subsection{Free Photon Action}
The photon action can be written as
    \begin{align}
    \act{\text{spin-1 kin}}
    =&
       \int_\Omega \dd^D x\,
\sqrt{-\det g}
      \Bigl(
       -\oh
       g^{\alpha\beta} g^{\gamma\delta}
       D_\alpha a_\gamma ( D_\beta a_\delta - D_\delta a_\beta)
       \Biggr)
      .
    \end{align}
If we choose the Lorenz gauge\footnote{Indeed it is
exactly the usual Lorenz gauge since locally the space is Minkowski.}
\begin{equation}
  D^\alpha a_\alpha
  =
  -\frac{1}{u} a_{v}
  -\du a_{v}
  -\dv a_{u}
  +\frac{1}{\Delta^2 u^2} \dz a_z
  + \eta^{ij} \partial_i a_j
  = 0
  \label{eq:Lorenz_gauge}
\end{equation}
and remember that covariant derivatives commute since we are locally flat,
the equations of motion read $(\square a)_\alpha=0$. Explicitly we have:
\begin{equation}
    \begin{split}
        ( \square a )_u
        & =
        \frac{1}{u^2} a_{v}
        -
        \frac{2}{\Delta^2 u^3} \dz a_z
        +
        \left[
            -
            2 \du \dv
            -
            \frac{1}{u} \dv
            +
            \frac{1}{\Delta^2 u^2} \dz^2
            +
            \eta^{ij} \partial_i \partial_j
        \right]
        a_u,
        \\
        ( \square a )_v
        & =
        \left[
            -
            2 \du \dv
            -
            \frac{1}{u} \dv
            +
            \frac{1}{\Delta^2 u^2} \dz^2
            +
            \eta^{ij} \partial_i \partial_j
        \right]
        a_v,
        \\
        ( \square a )_z
        & =
        -
        \frac{2}{u} \dz a_v
        +
        \left[
            -
            2 \du \dv
            +
            \frac{1}{u} \dv
            +
            \frac{1}{\Delta^2 u^2} \dz^2
            +
            \eta^{ij} \partial_i \partial_j
        \right]
        a_z,
        \\
        ( \square a )_i
        & =
        \left[
            -
            2 \du \dv
            -
            \frac{1}{u} \dv
            +
            \frac{1}{\Delta^2 u^2} \dz^2
            +
            \eta^{ij} \partial_i \partial_j
        \right]
        a_i
        .
    \end{split}
\end{equation}

As in the scalar case we are actually interested in solving
the eigenmodes problem $(\square a)_\alpha= r \,a_\alpha$.
We proceed hierarchically:
first we solve for
$a_v$ and $a_i$ whose equations are the same as the one for the scalar field,
then we insert the solutions as a source in the equation\footnote{Notice that inside the square brackets of the differential equation for $a_z$ there is a different sign for the term $\frac{1}{u} \dv$ with respect to the equation for the scalar field.} for $a_z$
and eventually we solve for $a_u$.
We get the solutions:
\begin{equation}
    \begin{split}
        \parallel\! {\tilde a}_{\kmkr\, \alpha}(u) \!\parallel
        \,=
        \begin{pmatrix}
            {\tilde a}_u
            \\
            {\tilde a}_v
            \\
            {\tilde a}_z
            \\
            {\tilde a}_i
        \end{pmatrix}
        & =
        \!\!\sum_{\underline{\alpha}
              \in 
              \{ \underline{u}, \underline{v}, \underline{z},\underline{i} \}
             }
         \!\!\cNE{\alpha}
         \parallel\! {\tilde a}^{\underline{\alpha}}_{\kmkr\, \alpha}(u) \!\parallel
        \\
        & =
        \cNE{u}
        \begin{pmatrix}
            1
            \\
            0
            \\
            0
            \\
            0
        \end{pmatrix}\,
        \tpkmkr(u)
        \\
        & +
        \cNE{v}
        \begin{pmatrix}
            \frac{i}{2 \kp u}
            +
            \frac{1}{2} \left( \frac{l}{\Delta \kp} \right)^2 \frac{1}{u^2}
            \\
            1
            \\
            \frac{l}{\kp}
            \\
            0
        \end{pmatrix}\,
        \tpkmkr(u)
        \\
        & +
        \cNE{z}
        \begin{pmatrix}
            \frac{l}{\Delta \kp \abs{u}}
            \\
            0
            \\
            \Delta \abs{u}
            \\
            0
        \end{pmatrix}\,
        \tpkmkr(u)
        \\
        & +
        \cNE{j}
        \begin{pmatrix}
            0
            \\
            0
            \\
            0
            \\
            \delta_{\underline{ij}}
        \end{pmatrix}\,
        \tpkmkr(u)
        ,
        \label{eq:Orbifold_spin1_pol}
    \end{split}
\end{equation}
  then we can expand the off-shell fields as
  \begin{equation}
    a_{\alpha}(u,v,z,\vec x)
    =\int \dd^{D-3} \vec k\, \int \dd r \int \dd \kp\,  \sum_{\mm\in\Z}
     \sum_{\underline{\alpha}
     \in \{ \underline{u}, \underline{v}, \underline{z},\underline{i}
     \} } \!\!\!\!
     \cNE{\alpha}\,
     {a}^{\underline{\alpha}}_{\kmkr\, \alpha}(u,v,z,\vec x) 
    ,
\end{equation}
  where ${a}^{\underline{\alpha}}_{\kmkr\, \alpha}(u,v,z,\vec x) =
    {\tilde a}^{\underline{\alpha}}_{\kmkr\, \alpha}(u)\, e^{i(k_+ v + l z + \vk
      \cdot \vec x)}$.

We can also compute the normalization as
\begin{equation}
\begin{split}
  (a_{(1)}, a_{(2)})
  &=
        \int \dd^{D-3} \vec x\,
      \int \dd u\, \int \dd v\, \int_0^{2\pi} \dd z\,
    | \Delta u|
    \\
  & 
    \times
    \Bigl(
    g^{\alpha\beta}
    a_{\kmkrN 1\, \alpha}\,  a_{\kmkrN 2\, \beta}
    \Bigr)
    \\
  &=
        \cEN{1} \circ \cEN{2} \,
    \\
  &\times
    \delta^{D-3}( \vkN{1} + \vkN{2})\,
    \delta( r{}_{(1)} - r{}_{(2)})\,
    \delta( \kpN{1} + \kpN{2})\,
    \delta_{\mmN{1} , -\mmN{2}}
    ,
\end{split}
\end{equation}
with\footnote{We use a shortened version of the polarizations $\cE$ for the sake of readability. Specifically we write $\cE_{(n)\, \underline{\alpha}} = \cE_{\kmkrN{n}\, \underline{\alpha}}$ thus hiding the understood dependence of the components of $\cE_{(n)}$ on the momenta.}
\begin{equation}
    \begin{split}
        \cE_{(1)} \circ \cE_{(2)}
        =
        -\cE_{(1)\, \underline{u}}~\cE_{(2)\, \underline{v}}
        -\cE_{(1)\, \underline{v}}~\cE_{(2)\, \underline{u}}
        +\cE_{(1)\, \underline{z}}~\cE_{(2)\, \underline{z}}
        + \eta^{\underline{ij}}\, \cE_{(1)\, \underline{i}}  \cE_{(2)\, \underline{j}}.
    \end{split}
\end{equation}
Finally the Lorenz gauge reads
\begin{equation}
     \eta^{ij} k_i \, \cNE{j} - \kp \cNE{u}
       - \frac{\vec k^2+r}{2 \kp} \cNE{v}
  =0
  ,
  \label{eq:explicit_orbifold_Lorenz}
\end{equation}
which does not impose any constraint on the transverse polarization
$\cNE{z}$,
and the photon kinetic term becomes 
\begin{align}
    \act{\text{spin-1 kin}}
    =&
      \int \dd^{D-3} \vec k\,
      \int \dd r
      \int \dd \kp\, 
      \sum_{\mm\in\Z}\,
      \oh
       r\,
       \cE_\kmkr\,
    \circ
       \cE_\kmkr^*
       .
  \end{align}
\subsection{Cubic Interaction}

With the definition of the d'Alembertian eigenmodes we can now examine the
cubic vertex which reads
  \begin{align}
    \act{\text{cubic}}
    =&
       \int_\Omega \dd^D x\,
       \sqrt{- \det g}
      \Bigl(
       -i\,e\,
       g^{\alpha\beta}
       a_\alpha
       (
      \phi^*\, \partial_\beta \phi
       -
       \partial_\beta \phi^* \,\phi
       )
       \Bigr)
       .
\end{align}
Its computation involves integrals such as
\begin{align}
    &
     \int \dd u\, \abs{ \Delta u}\,
     \left(\frac{\mm}{ u} \right)^2
     \prod_{i=1}^3 \tpkmkrN{i}
     \sim
     \int_{u\sim 0} \dd u\, 
     \left(\frac{l^2}{ |u|^{5/2} } \right)
     e^{-i
     \sum_{i=1}^3 \frac{ \mmN{i}^2 }{ 2 \Delta^2 \kpN {i}}
     \frac{1}{u}
      }
      ,
\end{align}
and
\begin{align}
     \int \dd u\, \abs{ \Delta u}\,
     \left(\frac{1}{ u} \right)
     \prod_{i=1}^3 \tpkmkrN{i}
     \sim
     \int_{u\sim 0} \dd u\, 
     \left(\frac{1}{ u |u|^{1/2} } \right)
     e^{-i
     \sum_{i=1}^3 \frac{ \mmN{i}^2 }{ 2 \Delta^2 \kpN {i}}
     \frac{1}{u}
      }
      ,
\end{align}
which can be interpreted as hints that the theory may be troublesome.
The first integral would diverge if the factor
$e^{i \frac{\cA}{u}}$ were equal $1$.
Fortunately it happens
when all $\mmN{*}=0$
but in this case the integral vanishes (if we set  $\mmN{*}=0$ before
its evaluation).
This however suggests that when all $\mmN *=0$, i.e. when the eigenfunctions are constant along the compact direction $z$, 
something is happening.
On the other side when at least one $\mm$ is different from zero we
have an integral such as
\begin{equation}
  \int_{u\sim 0} \dd u\, 
  \abs{u}^{-\nu}\, e^{i \frac{\cA}{u}}
  \sim
  \int_{t\sim \infty} \dd t\, 
  t^{\nu-2}\, e^{i \cA t}
  .
  \end{equation}
All $\mmN *$ are discrete but $\kpN *$ are not, therefore $\cA$ has an
isolated zero but otherwise it has continuous value and may be given a distributional meaning,
similar to a derivative of the $\delta$.

The second integral has again issues when all $\mmN{*}=0$ and
since it is not proportional to any $\mm$  as it stands it is
divergent unless we take a principal part regularization which may be
meaningful.

With all these warnings we can give anyhow meaning to the cubic terms
and we get\footnote{The notation $(2) \rightarrow (3)$ means that all previous terms inside the curly brackets appear again in exactly the same structure but with momenta of particle (3) in place of those of particle (2).}
\begin{align}
    \act{\text{cubic}}
  =&
     \prod_{i=1}^3
     \left[
     \int \dd^{D-3} \vkN i\, \dd \rN i\, \dd \kpN i
     \sum_{\mmN i}
     \right]\,
     (2\pi)^{D-1}
     \delta \left({\sum \vkN i}\right)\, \delta\left(\sum \kpN i \right)\,
     \nonumber\\
   \times&\, \delta_{\left(\sum \mmN i \right)}
     e \,
     (\cA_{\MINUSkmkrN 2})^* \cA_{\kmkrN 3}
\nonumber\\
   \times&\, \Biggl\{
     \cNEN u 1\, \kpN2\, \INT{3}{0}
     \nonumber\\
   +&\, \cNEN z 1\, \frac{\kpN 2 \mmN 1 - \mmN 2 \kpN 1}{\Delta \kpN 1}
     \MINT{3}{-1}
     \nonumber\\
   +&\, \cNEN v 1\,
     \Bigr[
     \frac{\vkN 2 ^2 +\rN 2}{2 \kpN 2} \INT{3}{0}
     +
     i
     \frac{\kpN 2}{ 2 \kpN 1} \INT{3}{-1}
     +
     \oh \frac{\kpN 2}{\Delta^2}
     \left( \frac{\mmN 1}{\kpN 1} - \frac{\mmN 2}{\kpN 2} \right)^2
     \INT{3}{-2}
     \Bigr]
     \nonumber\\
   -&\, \eta^{ij}\, \cNEN i 1\, k_{(2)_j}\, \INT{3}{0}\,
     -\Bigl( (2) \rightarrow (3) \Bigr)
     \Biggr\} 
     ,
       \label{eq:sQED_cubic_final}   
\end{align}
where we have defined also for future use
\begin{align}
  \INText{(1)\dots (N)}{\nu}
  =
  \INT{N}{\nu}
  &=
    \int_{-\infty}^{+\infty} \dd u\, \abs{\Delta u}\,u^\nu
     \prod_{i=1}^N \tpkmkrN{i}
      =
    \int_{-\infty}^{+\infty} \dd u\, \abs{\Delta u}\,u^\nu
     \prod_{i=1}^N \tponlyN{i} 
    ,
    \nonumber\\
  \MINT{N}{\nu}
  &=
    \int_{-\infty}^{+\infty} \dd u\, \abs{\Delta} \abs{u}^{\nu+1}
    \prod_{i=1}^N \tpkmkrN{i}
    ,
\end{align}
where 
$\tponlyN{i}=\tpkmkrN{i}$
and 
$\tponlyN{i}=\tpkmkrN{i}$
which will be used when not causing confusion.

\subsection{Quartic Interactions and Divergences}

In the previous section we have seen that the theory may have issues
when all $\mm=0$, i.e. with eigenfunctions independent of the compact direction $z$ because some integrals were on the verge of
diverging.
The divergence issue will appear in a clear and unavoidable way when
considering the quartic terms:
\begin{align}
  \act{\text{quartic}}
  &= \int_\Omega \dd^D x\,
\sqrt{- \det g}
    \left(
    e^2\, g^{\mu\nu}\, a_\mu a_\nu\, |\phi|^2
    -
    \frac{g_4}{4}
    \abs{\phi}^4
    \right)
,
  \end{align}
  which can be expressed using the modes as
  \begin{align}
    \act{\text{quartic}}
  =&
     \prod_{i=1}^4
     \left[
     \int \dd^{D-3} \vkN i\, \dd \kpN i\, \dd \rN i\,
     \sum_{\mmN i}
     \right]\,
     (2\pi)^{D-1}
     \delta\left( \sum \vkN i \right)\,
     \delta\left( \sum \kpN i \right)\, \delta_{\sum \mmN i,\, 0}
     \nonumber\\
     \Bigl\{
    e^2\,
    &(\cA_{\MINUSkmkrN 3})^* \cA_{\kmkrN 4}
\nonumber\\
   &
     \Bigl[
     (\cNEN{}{1} \circ \cNEN{}{2})\, \INT{4}{0}
     \nonumber\\
    &
     -
     i \oh
     \cNEN v 1 \cNEN v 2
     \left(
     \frac{1 }{\kpN 2}
     +
     \frac{1}{\kpN 1}
     \right)\,
     \INT{4}{-1}
     \nonumber\\
    &
      +\oh
      \frac{ \cNEN v 1 \cNEN v 2 }{\Delta^2}
      \left( \frac{\mmN 1}{\kpN 1} - \frac{\mmN 2}{\kpN 2} \right)^2
      \INT{4}{-2}
      \Bigr]
           \nonumber\\
      -
    \frac{g_4}{4} 
    &
    (\cA_{\MINUSkmkrN 1})^* (\cA_{\MINUSkmkrN 2})^* \cA_{\kmkrN 3}
      \cA_{\kmkrN 4}\,
      \INT{4}{0}
      \Bigr\}
      .
  \end{align}
  Now when setting $\mmN{*}=0$ all the surviving terms are divergent,
  explicitly
$\INT{4}{0}\sim \int d u\, |u|^{1-4\times\oh}$ and
$\INT{4}{-1}\sim \int d u\, |u|^{1-4 \times \oh} \frac{1}{u}$
since $\tilde \phi|_{\mm=0}\sim |u|^{-\oh}$.

Obviously higher order terms in the effective field theory will behave
even worse.
This makes the theory ill defined and the string theory which should
give this effective theory ill defined too.

\subsection{Failure of Obvious Divergence Regularizations}
\label{sec:saving}

From the discussion in the previous section it is clear that the
origin of the divergences is the sector $\mm=0$.
When $\mm=0$ the highest order singularity of the Fourier transformed
d'Alembertian equation vanishes. Explicitly we have:
    \begin{align}
      &\A \partial_u \tpkmkr+ \BB \tpkmkr=
        \nonumber\\
      &
      \A e^{-\int^u \frac{\BB}{\A} du}
      \partial_u\left[
      e^{+\int^u \frac{\BB}{\A} du}
      \tpkmkr
      \right]
      =0
      ,
    \end{align}
    with
    \begin{equation}
      \A=(-2i\, \kp),
      ~~~~
      \BB=
      (-\vec k^2-r)
      +
      (-i \kp) \frac{1}{u}
      +
      \frac{-\mm^2}{\Delta^2} \frac{1}{u^2}
      ,
    \end{equation}
and this in turn implies the absence
of the oscillating factor $e^{i \frac{\cA}{u} }$ when $l$ vanishes.
It follows that any deformation which makes the coefficient of the
highest order singularity not vanishing will do the trick.

The first and easiest possibility is to add a Wilson line along $z$,
i.e. $a= \theta d z$. This shifts $\mm \rightarrow \mm -e \theta$ and
regularizes the scalar QED.
Unfortunately this does not work for String Theory where
Wilson lines on $D25$ branes  are not felt by the neutral strings
starting and ending on the same brane. 
This happens because not all interactions involve commutators of the Chan-Paton factors which vanish for neutral strings. For instance the interaction among two tachyons and the first massive state involves an anti-commutator as we discuss later. 
The anti-commutators are present also in amplitudes of supersymmetric strings with massive states and therefore the issue is not solved by supersymmetry.

A second possibility is to think about higher derivative couplings to
curvature which is also natural in String Theory
If we regularize the metric in a minimal way as shown at the end of Section~\ref{sec:geometric_preliminaries_nbo}, we see that only $Ric_{u
  u}$ is non vanishing, therefore it would be natural to introduce
  \begin{align}
  S_{\mbox{higher R}}
  &= \int_\Omega \dd^D x\,
    \sqrt{- \det g}
    \left(
    \sum_{k\ge 1}
    \ap^{2 k-1}\,
    \prod_{j=1}^k g^{\mu_j\nu_j}\,g^{\rho_j\sigma_j}\,
    Ric_{\mu_j\rho_j}\,
    (\sum_{s=0}^{2 k} c_{k\, s} \partial^{2k-s}\phi^* \partial^s \phi)
    \right)
    \nonumber\\
  &= \int_\Omega \dd^D x\,
    \sqrt{- \det g}
    \left(
    \ap
    g^{\mu\nu}\,g^{\rho\sigma}\,
    Ric_{\mu\rho}\,
    (
    c_{1 2}
    \phi^* \partial^2_{\nu\sigma} \phi
    +
    c_{1 1}
    \partial_{\nu} \phi^* \partial_{\sigma} \phi
    +
    c_{1 0}
    \partial^2_{\nu\sigma} \phi^* \phi)
    \right)
,
  \end{align}
where $\ap$ has been introduced for dimensional reasons and in order
to have all $c$'s adimensional.
Since only $Ric_{u u}$ is non vanishing and it
depends only on $u$,
the regularized d'Alembertian eigenmode problem would now read
\begin{align}
  -2 \partial_u \partial_v \phi_r
  &
      -
    \frac{u}{u^2+\epsilon^2} \partial_v \phi_r
      +
    \frac{1}{\Delta^2 (u^2+ \epsilon^2)} \partial_z^2 \phi_r
    \nonumber\\
    &
    +
    \sum_{k\ge 1} \ap^{2k-1}
    C_k\,
    Ric_{u u}^k\,
    \partial_v^{2k} \phi
    +
      \partial_i^2 \phi_r
      -
      r \phi_r
      =0
      ,
\end{align}
with $C_k=  \sum_{s=0}^{2k} (-)^s c_{k\, s}$.
We can perform the usual Fourier transform and the function $B(u)$
becomes
\begin{equation}
\begin{split}
    \BB
    & =
    (-\vec k^2-r)
    +
    (-i \kp) \frac{u}{u^2+\epsilon^2}
    +
    \frac{-\mm^2}{\Delta^2} \frac{1}{u^2+\epsilon^2}
    \\
    &
    +
    \sum_{k\ge 1} \ap^{2k-1}
    C_k\,
    \left(\frac{\epsilon^2}{(u^2+\epsilon^2)^2}\right)^k
    (-i \kp)^{2k}
    .
\end{split}
\end{equation}
Then we examine what happens when $u=0$:
\begin{equation}
  B(0)\sim
      \frac{-\mm^2}{\Delta^2} \frac{1}{\epsilon^2}
    +
    \sum_{k\ge 1} \ap^{2k-1}
    C_k\,
    (-i \kp)^{2k}
    \frac{1}{\epsilon^ {2 k}}
    .
  \end{equation}
  Even though it looks as if it presents the possibility to cure the issue,
  unfortunately it is not so.
  If we consider $\ap$ and $\epsilon^2$ uncorrelated we lose
  predictability but
  if we consider $\ap\sim \epsilon^2$, as it is natural in String
  Theory,
  we do not solve the problem since 
 $  B(0)\sim
      \frac{-\mm^2}{\Delta^2} \frac{1}{\epsilon^2}
    +
    \sum_{k\ge 1} 
    C_k\,
    (-i \kp)^{2k}
    {\epsilon^ {2 k-2}}
    $
    and the curvature terms are not anymore singular.

\section{\NBO Eigenfunctions from Covering Space}
\label{sect:Eigenmodes_from_Covering}

In this section we recover the eigenfunctions from the covering
Minkowski space in order to elucidate the connection between the
polarizations in \NBO and in Minkowski.
Moreover we want to generalize the result to a symmetric two index
tensor which is the polarization of the first massive state to compute the two tachyons one massive
state in the next section and to show that it diverges.

\subsection{Spin 0 Wave Function from Minkowski space}
We start with the usual plane wave in flat space and we express it in
the new coordinates (we do not write the dependence on $\vec x$ since it is trivial)
\begin{align}
  \psi_{\kkk}(\xp,\xm,\xx)
  &=
    e^{i\left( \kp \xp + \km \xm + \kk \xx \right)}
    \nonumber\\
  =\psi_{\kkk}(u,v,z)
  &=
    e^{i\left[ \kp v
    +
    \frac{2 \kp \km - \kk^2}{2 \kp} u
    + \oh \Delta^2 \kp u \left( z+ \frac{\kk}{\Delta \kp}\right)^2
    \right]}
    .
  \end{align}
The corresponding wave function on the \NBO is obtained by
making it periodical in $z$.
This can be done in two ways either in $x$ coordinates or in $u v z$
ones.
The first way is more useful in deducing how the passage to the
orbifold makes the function depend on the equivalence class of momenta.
Implementing the projection on periodic $z$ functions  we get 
\begin{align}
  \Psi_{[\kkk]}([\xp,\xm,\xx])
  &=
  \sum_{n\in\Z}
    \psi_{\kkk}( \Ki{n }(\xp,\xm,\xx))
    \nonumber\\
    &=
  \sum_{n\in\Z}
    \psi_{\Ki{-n }( \kkk)}( \xp,\xm,\xx)
,
\end{align}
where we write $[\kkk]$ because the function depends on the equivalence
class of $\kkk$ only. The equivalence relation is given by
\begin{equation}
k=
\begin{pmatrix}
   \kp\\ \km\\ \kk
 \end{pmatrix}
  \equiv
  \Ki{-n } k =
  \begin{pmatrix}
  \kp 
  \\ 
  \km + n \hD \kk + \oh n^2 \hD^2 \kp
  \\
  \kk + n \hD \kp
  \end{pmatrix}
  ,
  \end{equation}
  and allows to choose a representative with
  \begin{equation}
    \left\{
      \begin{array}{l c}
        0\le \frac{\kk}{\Delta\, |\kp| }< 2 \pi & \kp\ne 0
        \\
        0\le \frac{\km}{\Delta\, |\kk| }< 2 \pi & \kp=0,\, \kk\ne 0
      \end{array}
      \right.
.
\end{equation}
  
If we perform the computation in $u v z$ coordinates we get
\begin{align}
  \Psi_{[\kkk]}(u,v,z)
  &=
  \sum_{n\in\Z}
  \psi_{\kkk}(u,v,z+2\pi\, n)
    \nonumber\\
  &=
  \sum_{n\in\Z}
    e^{i\left\{ \kp v
    +
    \frac{r}{2 \kp} u
    + \oh \hD^2 \kp u
    \left[ n +\frac{1}{2\pi} \left( z+ \frac{\kk}{\Delta \kp} \right)\right]^2
    \right\}}
    ,
  \end{align}
  with $r=2 \kp \km - \kk^2$
  and $Im(\kp u)>0$, i.e. $\kp u= |\kp u| e^{i \epsilon}$ and
  $\pi>\epsilon>0$.
  Notice that there is no separate dependence on $z$ and on $\frac{\kk}{\Delta \kp} $
  therefore one could fix the range $0\le z+\frac{\kk}{\Delta \kp}< 2\pi $.
  However this symmetry is broken when considering the photon eigenfunction.

  We can now use the Poisson resummation
  \begin{align}
    \sum_n e^{i a (n + b)^2}
    &= \int \dd s\, \delta_P(s) e^{i a (s+b)^2}
    =
      \frac{e^{-i  ( \frac{\pi}{4} + \oh arg(a)) } }{2 \sqrt{\pi | a
      |}}
      \sum_m e^{+ \frac{\pi^2 m^2}{i a}    + i 2\pi b m}
,
  \end{align}
  to finally get, reintroducing the other variables $\vk, \vex$ and
  setting therefore $r= 2 \kp \km -\kk^2-\vk^2$,
\begin{align}
  \Psi_{[\kkkk]}(u,v,z,\vec x)
  &=
    \sqrt{\frac{2}{\pi}}
    \frac{e^{-i \pi/4}}{\hD}
    \sum_\mm
    \left[
    \frac{1}{\sqrt{|\kp u|}}
    e^{i\left\{ \kp v
    +
    \mm z
    -
    \frac{\mm^2}{ 2 \Delta^2 \kp} \frac{1}{u}
    +
    \frac{r+ \vk^2}{2 \kp} u
    +
    \vk \cdot \vec x
    \right\}
    }
    \right]
    e^{ i \mm \frac{\kk}{ \Delta \kp} }
    \nonumber\\
  &=
    \cN
    \sum_l
\phi_\kmkr(u,v,z,\vec x)    
    e^{ i l \frac{\kk}{ \Delta \kp} }
~~~~\mbox{ when } \kp\ne 0
    ,
\label{eq:Psi_phi}
\end{align}
with
\begin{equation}
\cN=
        \sqrt{\frac{(2\pi)^D}{\pi \Delta}}
    \frac{e^{-i \pi/4}}{\pi }
  .
  \end{equation}
The fact that $\Psi$ depends only on the equivalence class
$[\kkkk]$ allows to restrict to $0\le \frac{\kk}{\Delta\, |\kp| }< 2
\pi$ so that we can invert the previous expression and get
\begin{align}
\phi_\kmkr(u,v,z,\vec x)    
  &=
    \frac{1}{\cN}
    \int_0^{2\pi \Delta |\kp|} \frac{\dd\kk}{ 2 \pi \Delta |\kp|}\,
    e^{ -i l \frac{\kk}{ \Delta \kp} }\,
  \Psi_{[\kkkk]}(u,v,z,\vec x)
  .
\end{align}

\subsection{Spin 1 Wave Function from Minkowski space}
We can repeat the steps of the previous section in the case of an electromagnetic
wave.
Again  we concentrate on $\xp, \xm$ and $\xx$ coordinates and
reinstate $\vec x $ at the end.
We start with the usual plane wave in flat space 
$\psi^{[1]}_{\kkk, \eee}$
and we express it in
both Minkowskian and orbifold coordinates.
We use the notation $\psi^{[1]}_{\kkk, \eee}$ to stress that it is the
eigenfunction and not the field which is obtained as
\begin{equation}
  A_\mu(x)\, d x^\mu
  =\int \dd^3 k\,
  \sum_{\epsilon }\psi^{[1]}_{\kkk, \eee}
  ,
\end{equation}
where the sum  is performed over $\epsilon$ which are 
independent and compatible with $k$.
The explicit expression for the eigenfunction with $\epsilon$ constant is
\footnote{We introduce the normalization factor $\cN$ in order to have a less cluttered relation between $\epsilon$ and $\cE$.}
\begin{align}
  &
\cN
  \psi^{[1]}_{\kkk, \eee}(\xp,\xm,\xx)
    (\epp d \xp + \epm d\xm + \epd d \xx)
    e^{i\left( \kp \xp + \km \xm + \kk \xx \right)}
    \nonumber\\
    &=
    \cN
  \psi^{[1]}_{\kkk, \eee}(u,v,z)
  =
    \left(
    \epu\, \dd u + \epz\, \dd z + \epv\, \dd v
    \right)
    e^{i\left[ \kp v
    +
    \frac{2 \kp \km - \kk^2}{2 \kp} u
    + \oh \Delta^2 \kp u \left( z+ \frac{\kk}{\Delta \kp}\right)^2
    \right]}
    ,
\end{align}
with
\begin{align}
  \epv &= \epp
         ,
         \nonumber\\
  \epu(z) &= \epm + (\Delta\, z) \epd + (\oh \Delta^2\, z^2) \epp
            ,
            \nonumber\\
  \epz(u,z) &= (\Delta\, u) (\epd + \Delta\, z\, \epp )
         .
\end{align}
Notice that we are not yet imposing any gauge
and also that
if $(\epp, \epm, \epd)$ are constant $(\epu, \epv, \epz)$ are
generically functions
but it is worth
stressing that $(\epu, \epv, \epz)$ are not the polarizations in the
orbifold which are anyhow constant, the fact that they depend on the
coordinates is simply the statement that not all eigenfunctions of the
vector d'Alembertian are equal.

Building the corresponding function on the orbifold amounts to summing
the images
\begin{align}
\cN
    \Psi^{[1]}_{[k,\, \epsilon]}([x])
  =&
    \sum_n    \epsilon \cdot ( \Ki{ -n } \dd x)~
     \psi_{k}( \Ki{ -n } x)
     =
     \sum_n   \Ki{ n } \epsilon \cdot \dd x~
\psi_{\Ki{ n } k}(x)
     ,
\end{align}
this expression makes clear that
under the action of the Killing vector $\epsilon$ transforms exactly
as the $k$ since it is induced by
$\epsilon \cdot \Ki{n} \dd x =  \Ki{-n } \epsilon \cdot \dd x
$, i.e.
\begin{equation}
\epsilon =
\begin{pmatrix}
   \epp\\ \epd \\ \epm
 \end{pmatrix}
  \equiv
  \Ki{-n } \epsilon =
  \begin{pmatrix}
  \epp 
  \\
  \epd + n \hD \epp
  \\ 
  \epm + n \hD \epd + \oh n^2 \hD^2 \epp
  \end{pmatrix}
  ,
  \end{equation}
however the pair 
$(k,\,\epsilon)$ transforms with the same $n$
since both are ``dual'' to $x$, i.e. their transformation rules are
dictated by the $x$.
Therefore there is only one equivalence class
$[k,\,\epsilon]$ and not two $[k]$, $[\epsilon]$.
Said differently,
a representative of the combined equivalence class is the one with
$0\le \kk< 2 \pi \Delta |\kp|$ when $\kp\ne 0$. 

We now proceed to find the eigenfunctions on the orbifold
in orbifold coordinates.
We notice that $\dd u, \dd v$ and $\dd z$ are
invariant and therefore their coefficients in $a$ are as well.
So we write
\begin{align}
\cN
    \Psi^{[1]}_{[k,\, \epsilon]}([x])
  =&
    \sum_n    \epsilon \cdot ( \Ki{ n } \dd x)\,
     \psi_{k}( \Ki{ n } x)
\nonumber\\
  =&
    \dd v\,
    \left[ \epp \sum_n \psi_{ k}( \Ki{ n } x) \right]
    +
    \dd z\, (\Delta u) 
    \left[
    \epd  \sum_n \psi_{ k}( \Ki{ n } x)
    +
    \epp \Delta   \sum_n (z + 2\pi n) \psi_{ k}(\Ki{ n }x)
    \right]
    \nonumber\\
  &+
    \dd u
    \left[
    \epm \sum_n \psi_{ k}(\Ki{ n }x)
    +
    \epd \Delta \sum_n  (z + 2\pi n) \psi_{ k}(\Ki{ n }x)
    +
    \oh
    \epp \Delta^2
    \sum_n (z + 2\pi n)^2 \psi_{ k}(\Ki{ n }x)
    \right]    
.
\end{align}
From direct computation we get\footnote{Notice that these expressions may be written using Hermite polynomials.}
\begin{align}
    &
    \sum_n (z + 2\pi n) \psi_{ k}(\Ki{ n }x)
    =
      \left(
          \frac{1}{i \Delta\, u}
      \frac{\partial}{\partial k_2}
      -
      \frac{\kk}{\Delta \kp}
      \right) \Psi_{[k]}([x])
      \nonumber\\
&
  \sum_n (z + 2\pi n)^2 \psi_{ k}(\Ki{ n }x)
    =
      \left(
          \frac{1}{i \Delta\, u}
      \frac{\partial}{\partial k_2}
      -
      \frac{\kk}{\Delta \kp}
      \right)^2 \Psi_{[k]}([x])
 .
\label{eq:sum_z_psi}
\end{align}
  Then it follows that
\begin{align}
\cN
  \Psi^{[1]}_{[k,\, \epsilon]}([x])
  =
  &
  \dd v\,
    \left[ \epp\,  \Psi_{[k]}([x])
    \vphantom{\frac{\partial}{\partial \kk}}
    \right]
    \nonumber\\
  &
    +
    \dd z\, (\Delta u)
    \left[
    \frac{\epd \kp -\epp \kk}{\kp}
     \Psi_{[k]}([x])
    + 
    \epp \,   \frac{-i}{u} \frac{\partial}{\partial \kk} \Psi_{[k]}([x])
    \right]
    \nonumber\\
  &+
    \dd u
    \Biggl[
    \left(
    \epm - \epd \frac{\kk}{\kp}
    + \oh \epp  \left( \frac{\kk}{\kp} \right)^2
    \right)
    \, \Psi_{[k]}([x])
    + 
    \frac{i}{2 u} \frac{\epp}{\kp }     \, \Psi_{[k]}([x])
    \nonumber\\
  &\phantom{d u [}
    +
    \frac{\epd \kp -\epp \kk}{\kp}
    \frac{-i}{ u} \frac{\partial}{\partial \kk} \Psi_{[k]}([x])
    +
    \oh
    \epp
    \frac{-1}{  u^2} \frac{\partial^2}{\partial \kk{}^2} \Psi_{[k]}([x])
    \Biggr]    
    ,
    \label{eq:a_uvz_from_covering}
\end{align}
where many coefficients of $\Psi$ or its derivatives  contain
$\kk$. 
They cannot be expressed using the quantum
numbers of the orbifold $\kmkr$ but are invariant on the orbifold and
therefore are new orbifold quantities which we can interpret as 
orbifold polarizations.
Using \eqref{eq:Psi_phi} we can finally write
\begin{align}
    \Psi^{[1]}_{[k,\, \epsilon]}([x])
  =
    \sum_l
  &  
\phi_\kmkr(u,v,z,\vec x)    
    e^{ i l \frac{\kk}{ \Delta \kp} }
    \Bigg\{
    \dd v \Bigl[ \epp \Bigr]
    \nonumber\\
  &+
    \dd z\,    (\Delta u)
    \Biggl[ 
    \frac{\epd \kp -\epp \kk}{\kp}
     +
    \epp \frac{1}{\Delta u} \frac{l}{\kp}
    \Biggr ]
    \nonumber\\
  &+
    \dd u
    \Biggl[
    \left(
    \epm - \epd \frac{\kk}{\kp}
    + \oh \epp  \left( \frac{\kk}{\kp} \right)^2
    \right)
       + 
    \frac{i}{2 u} \frac{\epp}{\kp }
    \nonumber\\
  &
 \phantom{   \dd u [ }
   +
    \frac{\epd \kp -\epp \kk}{\kp}
    \frac{1}{u} \frac{l}{\Delta  \kp}
    +
    \epp \frac{1}{2 u^2}
    \left(\frac{l}{ \Delta \kp} \right)^2
    \Biggr]
    \Bigg\}
    .
    \label{eq:spin1_from_covering}
\end{align}

If we compare with \eqref{eq:Orbifold_spin1_pol} we find
\begin{align}
  \cNE{v} &= \epp
            \nonumber\\
  \cNE{z} &= \sgn(u)
            \frac{\epd \kp -\epp \kk}{\kp}
            \nonumber\\
  \cNE{u} &=
            \epm - \epd \frac{\kk}{\kp}
            + \oh \epp  \left( \frac{\kk}{\kp} \right)^2
            ,
            \label{eq:eps_calE}
\end{align}
which implies that the true polarizations $(\epp, \epm, \epd)$ 
and
$\cNE{*}$ are constant as it turns out from direct computation.

A different way of reading the previous result is that the
polarizations on the orbifold are the coefficients of the highest power
of $u$.

We can also invert the previous relations to get
\begin{align}
  \epp
  &=
  \cNE{v} 
            \nonumber\\
  \epd
  &=
  \cNE{z} \sgn(u) +  \frac{\kk}{\kp} \cNE{v}
            \nonumber\\
  \epm &=
         \cNE u
         + \frac{\kk}{\kp} \cNE{z} \sgn(u)
          + \oh \left( \frac{\kk}{\kp} \right)^2 \cNE{v} 
            ,
            \label{eq:calE_eps}
\end{align}
and use them in Lorenz gauge $k \cdot \epsilon=0$ in order to get the
expression of Lorenz gauge with orbifold polarizations.
If the definition of orbifold polarizations is right the result cannot
depend on $\kk$ since $\kk$ is not a quantum number of orbifold
eigenfunctions.
Taking in account $\km = \frac{\vk^2+ \kk^2 + r}{2 \kp}$ in $k \cdot
\epsilon=0$
we get
exactly the expression for the Lorenz gauge for orbifold polarizations
\eqref{eq:Lorenz_gauge}.


\subsection{Tensor Wave Function from Minkowski space}
Once again, we can use the analysis of the previous section in the case of a
second order symmetric tensor wave function.
Again we suppress the dependence on $\vec x$ and $\vk$ with a caveat: the Minkowskian polarizations $\Ss + i$, $\Ss - i$ and $\Ss 2 i$ do transform non trivially, therefore we give the full expressions in Appendix~\ref{app:NO_tensor_wave} even if these components contribute in a somewhat trivial way since they behave effectively as a vector of the orbifold.

We start with the usual wave in flat space and we express either in
the Minkowskian coordinates
\begin{alignat}{4}
\cN
  \psi^{[2]}_{k\, S}(\xp,\xm,\xx)
  &= S_{\mu \nu}\, \psi_k(x)\, \dd x^\mu\, \dd x^\nu
  \nonumber\\
  &=
    (
    \Ss + +\, \dd \xp\, \dd \xp
    &
    +
    2 \Ss + 2\, \dd \xp\, \dd \xx
    &
    +
    2 \Ss + -\, \dd \xp\, \dd \xm
    \nonumber\\
    &
    &
    +
    2 \Ss 2 2\, \dd \xx\, \dd \xx
    &
    +
    2 \Ss 2 -\, \dd \xx\, \dd \xm
    \nonumber\\
    &
    &
    &
    +
    2 \Ss - -\, \dd \xm\, \dd \xm
    )
    e^{i\left( \kp \xp + \km \xm + \kk \xx \right)}
    \nonumber\\
      ,
\end{alignat}
or in orbifold coordinates
\begin{align}
\cN
  \psi^{[2]}_{k\, S}(x)
  =& S_{\alpha \beta}\, \psi_k(x)\, \dd x^\alpha\, \dd x^\beta 
    \nonumber\\
  =&
  \Bigl\{
  (\dd v)^2\,
  [\Ss ++]
  \nonumber\\
  &
  +
  \dd v\, \dd z\,\Delta u
  [ 2 \Ss+2 
  + \Ss++ \Delta z ]
  \nonumber\\
  &
  +
  \dd v\, \dd u\,
  [ 2 \Ss+-
  +  2 \Ss+2 \Delta z
  + \Ss++ \Delta^2 z^2 ]
  \nonumber\\
  &
  +
  \dd z^2\,\Delta^2 u^2\,
  [ \Ss22 
  + 2 \Ss+2 \Delta  z
  + \Ss++ \Delta^2 z^2 ]
  \nonumber\\
  &
  +
  \dd z\, \dd v\, \Delta u\,
  [ 2 \Ss-2 
  + 2 (\Ss22 + \Ss+- ) \Delta  z
  + 3 \Ss+2 \Delta^2  z^2
  + \Ss++ \Delta^3  z^3
  ]
  \nonumber\\
  &
  +
  \dd u^2\,
  [
  \Ss--
  + 2 \Ss-2 \Delta z
  + (\Ss22 + \Ss+-) \Delta^2 z^2
  + \Ss+2 \Delta^3 z^3
  + \frac{1}{4} \Ss++ \Delta^4 z^4
  ]
  \Bigr\}
    \nonumber\\
   &\times
     e^{i\left[ \kp v
    +
    \frac{2 \kp \km - \kk^2}{2 \kp} u
    + \oh \Delta^2 \kp u \left( z+ \frac{\kk}{\Delta \kp}\right)^2
    \right]}
.
\end{align}
Now we define the tensor on the orbifold as a sum over all images as
\begin{align}
\cN
  \Psi^{[2]}_{[k\, S]}([x])
  &=
    \sum_n    ( \Ki{ n }d x) \cdot S \cdot ( \Ki{ n } \dd x)~
    \psi_{k}( \Ki{ n } x)
    \nonumber\\
  &=
    \sum_n    \dd x \cdot ( \Ki{ - n }S ) \cdot \dd x~
    \psi_{ \Ki{ -n } k}( x)
    .
  \end{align}
  In the last line we have defined the induced action of the Killing vector on
  $(k, S)$ which can be explicitely written as
  \begin{align}
    \Ki{-n}
    \left(
    \begin{array}{c}
      S_{ + + }\\
      S_{ + 2 }\\
      S_{ + - }\\
      S_{ 2 2 }\\
      S_{ 2 - }\\
      S_{ - - }
    \end{array}
    \right)
    =
    \left(
    \begin{array}{c}
      S_{ + + }\\
      S_{ + 2 } + n \Delta S_{ + + }\\
      S_{ + - } + n \Delta S_{ + 2 } + \oh n^2 \Delta^2 S_{ + + }\\
      S_{ 2 2 } + 2 n \Delta S_{ + 2 } + n^2 \Delta^2 S_{ + + }\\
      S_{ 2 - } + n \Delta (S_{ 2 2 } +S_{ + - })
      + \frac{3}{2} n^2 \Delta^2 S_{ + 2 } + \oh n^3 \Delta^3 S_{ + + }\\
      S_{ - - } + 2 n \Delta S_{ - 2 }
      + n^2 \Delta^2 (S_{ 2 2 } + S_{ + - } ) + n^3 \Delta^3 S_{ + 2 }
      + \frac{1}{4} n^4 \Delta^4 S_{ + + }
    \end{array}
    \right)
.
    \end{align}
  
  In orbifold coordinates
  to compute the tensor on the orbifold simply
  amounts to sum over all the shifts
  $z \rightarrow (z+2\pi n)$ and the use of the generalization of
  \eqref{eq:sum_z_psi}, i.e. to substitute
  $(\Delta\,z)^j \psi_k \rightarrow
  \left(
    \frac{1}{i  u} \frac{\partial}{\partial \kk}
- \frac{ \kk}{ \Delta \kp}
  \right)^j
  \Psi_{[k]}([x])$.
  When expressing all in the $\phi$ basis
  this last step is equivalent to
  $(\Delta\, z)^j \psi_k \rightarrow
  \left( \frac{l}{\Delta\, u\, \kp}  \right)^j
  + \dots
  $.
  We identify the basic polaritazions on the orbifold by
  considering the highest power in $u$ and get
\newcommand{\RRatio}{ K }
  \begin{align}
    \cS_{u\,u}
    &=
      \frac{1}{4}{{\RRatio^4\,S_{+\,+}}}
      +\RRatio^2\,S_{+\,-}
      -\RRatio^3\,S_{+\,2}
      +S_{-\,-}
      -2\,\RRatio\,S_{-\,2}
      +S_{2\,2}\,\RRatio^2
\nonumber\\
    \cS_{u\,v}
    &=
      \oh {{\RRatio^2\,S_{+\,+}}}
      +S_{+\,-}
      -\RRatio\,S_{+\,2}
      \nonumber\\
    \cS_{u\,z}
    &=
      - \oh {{\RRatio^3\,S_{+\,+}}}
      -\RRatio\,S_{+\,-}
      +\frac{3}{2} {{\RRatio^2\,S_{+\,2}}}
      +S_{-\,2}
      -\RRatio\, S_{2\,2}
\nonumber\\
    \cS_{v\,v}
    &=
      S_{+\,+}
\nonumber\\
    \cS_{v\,z}
    &=
      S_{+\,2}-\RRatio\,S_{+\,+}
\nonumber\\
    \cS_{z\,z}
    &=
      \RRatio^2\,S_{+\,+}-2\,\RRatio\,S_{+\,2}+S_{2\,2}
.
  \end{align}
with $\RRatio= \frac{\kk}{ \kp}$.
The previous equations can be inverted into
\begin{align}
S_{-\,-}
&=
          \RRatio^2\,\left(\cS_{z\,z}+\cS_{u\,v}\right)
          +\RRatio^3\,\cS_{v\,z}
          +\frac{1}{4} \RRatio^4\,\cS_{v\,v}
          +2\,\RRatio\,\cS_{u\,z}
          +\cS_{u\,u}
\nonumber\\
S_{+\,-}
&=
          \RRatio\,\cS_{v\,z}
          +\oh \RRatio^2\,\cS_{v\,v}
          +\cS_{u\,v}
\nonumber\\
S_{-\,2}
&=
          \RRatio\,\left(\cS_{z\,z}+\cS_{u\,v}\right)
          +\frac{3}{2}\,\RRatio^2\,\cS_{v\,z}
          +\oh \RRatio^3\,\cS_{v\,v}
          +\cS_{u\,z}
\nonumber\\
S_{+\,+}
&=
\cS_{v\,v}
\nonumber\\
S_{+\,2}
&=
\cS_{v\,z}+\RRatio\,\cS_{v\,v}
\nonumber\\
S_{2\,2}
&=
\cS_{z\,z}+2\,\RRatio\,\cS_{v\,z}+\RRatio^2\,\cS_{v\,v}
.
\end{align}
Since we plan to use the previous quantities in the case of the first
massive string state we compute the relevant quantities: the trace
\begin{equation}
  \tr(S)=\cS_{z\,z}-2\,\cS_{u\,v}
\end{equation}
and the transversality conditions
\newcommand{\MMr}{ (r+\vk^2) }
\begin{align}
  %
  trans ~\cS_{v}
  =&
  (k\cdot S)_{+}
 =    
    -\frac{\MMr}{2\, \kp}\,
    \cS_{v\,v}
    -k_{+}\,\cS_{u\,v},
    \nonumber\\
    %
  trans ~\cS_{z}
  =&
     (k\cdot S)_{2}
     -\RRatio  (k\cdot S)_{+}
=
     -\frac{\MMr}{2\, \kp}\,
     \cS_{v\,z}
     -k_{+}\,\cS_{u\,z},
\nonumber\\
  trans ~\cS_{u}
  =&
     (k\cdot S)_{-}
     -\RRatio  (k\cdot S)_{2}
     + \oh \RRatio^2  (k\cdot S)_{+}
   =
     -\frac{\MMr}{2\, \kp}\,
     \cS_{u\,v}
     -k_{+}\,\cS_{u\,u}
     .
\end{align}
where we used $k_-= (r+\vk^2+k_2^2)/(2 k_+)$.
These conditions correctly do no depend on $\RRatio$ since $\kk$ is
not an orbifold quantum number.

The final expression for the orbifold symmetric tensor is
  \begin{align}
    \Psi^{[2]}_{[k,\, S]}\left([x]\right)
    &
    =
    \sum_l
\phi_\kmkr(u,v,z,\vec x)    
    e^{ i l \frac{\kk}{ \Delta \kp} }
\nonumber\\
%
    &
      \Big\{
      (\dd v)^2\,
      [\cS_{v v} ]
       \nonumber\\
%
    &
      +
2
      \Delta\, u\,
      \dd v\, \dd z\,
      \Bigl[
      \cS_{v\,z}
      +
      \left(
      \frac{L \cS_{v\,v}}{\Delta}
      \right)
      \frac{1}{u}
      \Bigr]
   \nonumber\\
%
  &
    +
    2
  \dd v\, \dd u\,
\Bigl[
\cS_{u\,v}
+
\left(
\frac{L\,\cS_{v\,z}}{\Delta}+\frac{i\,\cS_{v\,v}}{2\,k_{+}}
\right)
\frac{1}{u}
+
\left(
\frac{L^2\,\cS_{v\,v}}{2\,\Delta^2}
\right)
\frac{1}{u^2}
\Bigr]
    \nonumber\\
    %
  &
  +
    (\Delta\, u)^2
  \dd z^2\,
    \Bigl[
    \cS_{z\,z}
+
\left(
\frac{2\,L\,\cS_{v\,z}}{\Delta}+\frac{i\,\cS_{v\,v}}{k_{+}}
\right)
    \frac{1}{u}
    +
\left(
\frac{L^2\,\cS_{v\,v}}{\Delta^2}
\right)
\frac{1}{u^2}
    \Bigr]
    \nonumber\\
    %
  &
    +
    2
    \Delta u\,
  \dd z\, \dd u\,
    \Bigl[
    \cS_{u\,z}
+
\left(
    \frac{L\,\cS_{z\,z}}{\Delta}
    +\frac{3\,i\,\cS_{v\,z}}{2\,k_{+}}+\frac{L\,\cS_{u\,v}}{\Delta}
\right)
    \frac{1}{u}
    +
\left(
    \frac{3\,L^2\,\cS_{v\,z}}{2\,\Delta^2}
    +\frac{3\,i\,L\,\cS_{v\,v}}{2\,\Delta\,k_{+}}
\right)
    \frac{1}{u^2}
    \nonumber\\
    &
    \phantom{+\Delta u\,d z\, d v\,}
    +
\left(
\frac{L^3\,\cS_{v\,v}}{2\,\Delta^3}
\right)
\frac{1}{u^3}
    \Bigr]
      \nonumber\\
      %
  &
  +
  \dd u^2\,
  %
    \Bigl[
\cS_{u\,u}
+
\left(
\frac{i\,\cS_{z\,z}}{k_{+}}+\frac{2\,L\,\cS_{u\,z}}{\Delta}+\frac{i\,\cS_{u\,v}}{k_{+}}
\right)
    \frac{1}{u}
    +
\left(
    \frac{L^2\,\cS_{z\,z}}{\Delta^2}+\frac{3\,i\,L\,\cS_{v\,z}}{\Delta\,k_{+}}
    -\frac{3\,\cS_{v\,v}}{4\,k_{+}^2}+\frac{L^2\,\cS_{u\,v}}{\Delta^2}
\right)
    \frac{1}{u^2}
    \nonumber\\
    &
      \phantom{  +  d u^2\, }
    +
\left(
\frac{L^3\,\cS_{v\,z}}{\Delta^3}+\frac{3\,i\,L^2\,\cS_{v\,v}}{2\,\Delta^2\,k_{+}}
\right)
      \frac{1}{u^3}
      +
\left(
\frac{L^4 \cS_{v\,v}}{4 \Delta^4}
\right)
\frac{1}{u^4}
    \Bigr]
    \Bigr\}
    ,
  \end{align}
  with $L=\frac{l}{\kp}$.

\section{Overlap of Wave Functions and Their Derivatives}
\label{sec:overlap}
In this section we compute overlaps of wave functions and give their
expressions both using integrals
over the eigenfunctions and sum of products of $\delta$.
The latter is the expression which is naturally obtained
by computing tree level string amplitudes on the orbifold
when one starts with Minkowski amplitudes and adds the amplitudes due
to images.
This is equivalent to compute emission vertices on the orbifold and
then compute their correlations since this amounts to transferring the sum over the spacetime images to the sum of the polarizations images.
We show this carefully in the next section.
We consider also when and if they diverge.
Finally we use the wording wave function for the functions on Minkowski space because we do not assume any constraint on polarizations.

\subsection{Overlaps Without Derivatives}
Let us start with the simplest case of the overlap of $N$ scalar wave
function.
We compute the overlap of orbifold wave functions and then we
re-express it as sum of images of the corresponding Minkowski overlap
thus establishing a dictionary between Minkowski and orbifold overlaps.
Explicitely we consider the following overlap where all the polarizations $\cA_{(i)}$ have been set to one
\begin{align}
I^{(N)}=
  &
  \int_{\Omega} \dd^3x\,
  \sqrt{-\det g}\,
  \prod_{i=1}^N   \Psi_{[\kkkN{i}]}([\xp,\xm,\xx]))
  \nonumber\\
  &=
     \int_{\R^{1,2}} \dd^3x\,
     \sqrt{-\det g}~
     \psi_{\kkkN{1}}(\xp,\xm,\xx))
     \prod_{i=2}^N  \sum_{m_{(i)}\in\Z} 
     \psi_{\kkkN{i}}( \Ki{m_{(i)} }(\xp,\xm,\xx) )      
  \nonumber\\
  &=    
     \int_{\R^{1,2}} \dd^3x\,
     \sqrt{-\det g}~
     \psi_{\kkkN{1}}(\xp,\xm,\xx))
     \prod_{i=2}^N  \sum_{m_{(i)}\in\Z} 
    \psi_{ \Ki{m_{(i)} }(\kkkN{i}) }( \xp,\xm,\xx )
    \nonumber\\
  &=
    (2\pi)^3 \delta(\sum_i \kpN{i})\,
     \prod_{i=2}^N  \sum_{m_{(i)}\in\Z} 
    \delta\left(\sum_i \Ki{m_{(i)} } \kkN{i} \right)\,
    \delta\left(\sum_i \Ki{m_{(i)} } \kmN{i} \right)\,
    \Bigg|_{m_{(1)} = 0}
,
\end{align}
where $\Omega=\R^{1,2}/\Gamma$ is the orbifold identified with the
fundamental region of $\R^{1,2}/\Gamma$.
We used the unfolding
trick to rewrite the integral as an integral over $R^{1,2}$
thus dropping the sum over the images of particle $(1)$.
Then we moved the action of the Killing vector from $x$ to $k$
and finally we used the usual $\delta$ definition.
The previous integral can be expressed explicitely as
\begin{align}
  I^{(N)}=
    &
    \cN^N
     \sum_{ \{\mmN{i}\} \in\Z^N } 
    e^{i \sum_{i=1}^N \mmN{i} \frac{\kkN{i} }{ \Delta \kpN{i} } }
     \int_{\Omega} \dd^3x\,
     \sqrt{-\det g}\,
    \prod_{i=1}^N
    \pkmrN{i}([x]))
    \nonumber\\
  &=
    \cN^N
     \sum_{ \{\mmN{i}\} \in\Z^N } 
     e^{i \sum_{i=1}^N \mmN{i} \frac{\kkN{i} }{ \Delta \kpN{i} } }\,
     (2\pi)^2
     \delta\left( \sum \kpN{i} \right)
     \delta_{\sum \mmN{i}}\,
\INT{N}{0}
    ,
  \end{align}
  from which we can reexpress  the overlap of the wave functions using
  integrals over the infinite sum $\delta^2$ as
  \begin{align}
      \int_{\Omega} \dd^3x\,
&&&
    \prod_{i=1}^N   
    \pkmrN{i}\left( [x] \right))
    =
    \frac{1}{\cN^N}
    \prod_{i=1}^N   
    \int_0^{2\pi \Delta |\kpN{i}|}
    \frac{\dd\kkN{i}}{ 2 \pi \Delta |\kpN{i}|}\,
    e^{ -i \mmN{i} \frac{\kkN{i}}{ \Delta \kp{i}} }\,
      I^{(N)}
\nonumber\\    
    =&&&
    (2\pi)^3 \delta\left(\sum_i \kpN{i} \right)\,
    \frac{1}{\cN^N}
    \prod_{i=1}^N   
    \int_0^{2\pi \Delta |\kpN{i}|}
    \frac{\dd\kkN{i}}{ 2 \pi \Delta |\kpN{i}|}\,
       e^{ -i \mmN{i} \frac{\kkN{i}}{ \Delta \kpN{i}} }\,
       \nonumber\\
    &&&
       \prod_{j=2}^N  \sum_{m_{(j)}\in\Z} 
    \delta\left( \sum_j \Ki{m_{(j)} } \kkN{j} \right)\,
    \delta\left( \sum_j \Ki{m_{(j)} } \kmN{j} \right)\,
.
  \end{align}

In particular it follows from the explicit expression of $\INT{n}{0}$
that all overlaps $I^{(N)}$ for $N\ge 4$ are
infinite. 

Is there any intuitive reason for the divergence of the overlapping?
We are summing over infinite distributions with accumulation points of their support.
Nevertheless the existence of the accumulation point is not sufficient
since the three scalars overlap, i.e. the three tachyons amplitude converges: 
the coefficients of the
deltas matter, too, and the convergence must be verified. 

\subsection{An Overlap With One Derivative}
Since we will also compute the amplitude involving two tachyons and one photon, as a preliminary step we consider
the overlap in Minkowski space
\begin{equation}
J_{Mink}=     i\,      (\epN{1}\cdot \kkN{2})\,
  (2\pi)^3 \delta\left( \sum_i \kpN{i} \right)\,
    \delta\left( \sum_i  \kkN{i} \right)\,
     \delta\left( \sum_i \kmN{i} \right)\,
     .
 \end{equation}
 Applying the recipe of summing over momentum and polarizations
 images of all but one particle,
 we can produce an expression which depends on equivalence
classes  as
\begin{equation}
    \begin{split}
        & J([\kN{1}, \epN{1}],\,  [\kN{2}],\, [\kN{3}])
        =
        i\, (2\pi)^3 \delta\left( \sum_i \kpN{i} \right)\,
        \\
        & \times
        \sum_{\{ m_{(i)} \} \in\Z^3}
        \delta_{ m_{(1)},1}\,
        (\Ki{m_{(1)} }\epN{1} \cdot \Ki{m_{(2)} } \kkN{2})\,
        \delta\left( \sum_i \Ki{m_{(i)} } \kkN{i} \right)\,
        \delta\left( \sum_i \Ki{m_{(i)} } \kmN{i} \right).
    \end{split}
    \label{eq:Spin_001_overlap_from_covering}
\end{equation}
This expression depends only on equivalence classes,
  for example under $(\kN{1},\epN{1}) \rightarrow \Ki{s}(\kN{1},\epN{1})$,
  we can use $\Ki{s} a \cdot b = a \cdot \Ki{-s} b$ and the invariance
  of deltas $\delta^3(\Ki{s}a)=\delta^3(a)$ to demonstrate it.

  Now it is not difficult to show that the previous expression can be
  written as
  \begin{align}
    J
    &=
      \int_\Omega \dd^3x\,
      \eta^{\mu\nu}\,
      \Psi^{[1]}_{[\kN{1}, \epN{1}]\, \mu}([x])\,
      \partial_\nu \Psi_{ [\kN{2}] }([x])\,
      \Psi_{ [\kN{3}] }([x])
 \end{align}
 where we performed the unfolding using
 $a_{[\kN{1}, \epN{1}]\,\mu}([x])$.
 Obviously we can perform the unfolding using whichever other field and
this amount to keep the corresponding $m_{(i)}$ fixed in place of $m_{(1)}$.

Notice that the previous expression is invariant despite the fact that the
derivatives $\partial_\mu$ are not well defined on the orbifold since not invariant and would hamper the use of the unfolding trick.
The expression is invariant because $\Psi^{[1]}_\mu$ is not invariant too and
compensates.

We can then evaluate the previous expression with Minkowskian polarizations using \eqref{eq:spin1_from_covering} which is nothing else but a rearrangement of terms of \eqref{eq:Spin_001_overlap_from_covering}  to write
\begin{alignat}{3}
  J
  =
  &
     i\,
    \cN^2
  \sum_{ \{\mmN{i}\} \in\Z^3 }
  &&
    e^{i \sum_{i=1}^3 \mmN{i} \frac{\kkN{i} }{ \Delta \kpN{i} } }\,
     (2\pi)^2
     \delta\left( \sum \kpN{i} \right)
     \delta_{\sum \mmN{i}}
     \nonumber\\
  &    
  &&
  \times
     \int_{\Omega} \dd^3x\,
    \prod_{i=1}^3
    \pkmrN{i}([x]))
    \Big\{
    \eppN{1}\,
    \left[
    +
    \frac{i}{2 u}
    +
    \frac{\mmN{2}^2}{  \kpN{2}} \frac{1}{2 \Delta^2\, u^2}
    +
    \frac{\rN{2}}{2 \kpN{2}}
    \right]
    \nonumber\\
  &&&+
    \frac{1}{\Delta\, u}
    \left[ \epdN{1}\,
      + \frac{1}{\Delta u} \eppN{1} \frac{\mmN{1}}{\kpN{1}} \right]\,
    \mmN{2}
    \nonumber\\
   &&&+
    \left[\epmN{1} + \epdN{1} \frac{1}{\Delta u} \frac{\mmN{1}}{\kpN{1}}
    + \eppN{1} \frac{1}{2 (\Delta u)^2} \frac{\mmN{1}^2}{\kpN{1}^2}
  \right]\,
        \kpN{2}
        \Big\}
        .
    \label{eq:divergence_overlap_spin1}
\end{alignat}
Possible divergences come when $\mm=0$ because the absence of the factor
$e^{i \frac{A}{u} }$,
however all explicit factor $\frac{1}{u}$ come always with $\mm$ therefore
when $\mm=0$ they do not give any contribution.
A divergence when $\mm=0$ comes actually only 
from the contribution of the first line
$\partial_u\phi|_{\mm=0} =-\frac{1}{2u} \phi|_{\mm=0}$
but this cancels in scalar QED or with abelian tachyons because we have to subtract the contribution obtained exchanging $(2)$ and $(3)$. 
Because of color factors it does not cancel 
when considering the non abelian case unless one
uses a kind of principal part prescription
since replacing
$\int_{-|a|}^{|b|} \dd u\, \frac{\sgn(u)}{|u|^{3/2}}$
with
$\lim\limits_{\delta\rightarrow 0}
\left[\int_{-|a|}^{-|\delta|}+\int _{-|\delta|} ^{|b|} \right]
\dd u\, \frac{\sgn(u)}{|u|^{3/2}}$
gives a  finite result.

\subsection{An Overlap With Two Derivatives}
  We can generalize the previous expressions to more general cases.
  Since we use the results from Section~\ref{sect:Eigenmodes_from_Covering} we miss some non trivial contributions from polarizations like $\cS_{ v i}$.
  These contributions do not alter the discussion. However for completeness we give the lengthy full expression in Appendix~\ref{app:NO_full_TTS}.
  
  Having in mind the amplitudes with two tachyons and one massive state,
  we can consider an expression like 
  \begin{align}
    K
    &=
      \int_\Omega \dd^3x\,
      \sqrt{-\det g}\,
      \eta^{\mu\nu}\,\eta^{\rho\sigma}\,
      \Psi^{[2]}_{[\kN{3}, \SsN{3}]\, \mu\rho}([x])\,
      \partial^2_{\nu\sigma} \Psi_{ [\kN{2}] }([x])\,
      \Psi_{ [\kN{1}] }([x])
      ,
  \end{align}
  in Minkowskian coordinates or
\begin{align}
  K
  =
  \int_\Omega \dd^3x\,
  \sqrt{-\det g}~
  &
      g^{\alpha\beta}\,g^{\gamma\delta}\,
      \Psi^{[2]}_{[\kN{3}, \SsN{3}]\, \alpha\gamma}([x])\,
      D_\beta \partial_{\delta} \Psi_{ [\kN{2}] }([x])\,
      \Psi_{ [\kN{1}] }([x])
\end{align}
in orbifold coordinates where we need to use covariant derivatives.
Using the unfolding trick over $(3)$ we get
\begin{align}
  K=
  &
    (2\pi)^3 \delta\left( \sum_i \kpN{i} \right)\,
     \prod_{i=2}^N  \sum_{m_{(i)}\in\Z}
  S_{(3) \mu \rho}\,
  (\Ki{m_{(2)} } \kkN{2})^\mu
    (\Ki{m_{(2)} } \kkN{2})^\rho\,
    \nonumber\\
  &\times
    \delta\left( \sum_i \Ki{m_{(i)} } \kkN{i} \right)\,
    \delta\left( \sum_i \Ki{m_{(i)} } \kmN{i} \right)
    .
    \label{eq:Spin_002_overlap_from_covering}
\end{align}
Explicitly in orbifold coordinates we can write
\begin{align}
  K
  =
  \int_\Omega \dd^3x\,
  &
  \sqrt{-\det g}
    \Bigl[
    +
    \Psi^{[2]}_{[\kN{3}, \SsN{3}]\, u u }\,
    \partial_v^2 \Psi_{ [\kN{2}] }
    \nonumber\\
  &
    -
  2\frac{1}{ (\Delta u)^2 }
    \Psi^{[2]}_{[\kN{3}, \SsN{3}]\, u z }\,
    \partial_v \partial_z \Psi_{ [\kN{2}] }
    \nonumber\\
  &
    +
    2
    \Psi^{[2]}_{[\kN{3}, \SsN{3}]\, u v }\,
    \partial_v \partial_u \Psi_{ [\kN{2}] }
    \nonumber\\
  &
    +
  \frac{1}{ (\Delta u)^4 }
    \Psi^{[2]}_{[\kN{3}, \SsN{3}]\, z z }\,
    ( \partial_z^2 \Psi_{ [\kN{2}] } - \Delta^2 u \partial_v \Psi_{ [\kN{2}] }) 
    \nonumber\\
  &
    -2
  \frac{1}{ (\Delta u)^2 }
    \Psi^{[2]}_{[\kN{3}, \SsN{3}]\, z v }\,
    ( \partial_z \partial_u \Psi_{ [\kN{2}] } - \frac{1}{u} \partial_z \Psi_{ [\kN{2}] }) 
    \nonumber\\
&
  +
    \Psi^{[2]}_{[\kN{3}, \SsN{3}]\, v v }\,
    \partial_u^2 \Psi_{ [\kN{2}] }
  \Bigr]
                  \Psi_{ [\kN{1}] }
                  .
\end{align}
Keeping the terms which do not vanish when all $\mm=0$
and considering only the leading order in $\frac{1}{u}$
we get
\begin{align}
  K\sim
\int \dd u\, |u|\,
  &
\frac{3}{4} \frac{(\kpN 2 +\kpN 3)^2}{\kpN{3} ^2}
    \,\cS_{v v (3)}\,
    \frac{1}{u^2}
   \prod_{i=1}^3 \phi_{ (i) } \Big|_{\mmN{*}=0}  
    ,
    \label{eq:divergence_overlap_spin2}
\end{align}
which is divergent as $\frac{1}{|u|^{5/2}}$.

\section{String Three Points Amplitudes With One Massive State}
\label{sect:NO3ptsMassive}
In this section we consider string amplitudes including string massive states.
They are obtained using the inheritance principle and therefore they
are connected to the integrals and relations derived in Section~\ref{sec:overlap}.

In particular we want to use the inheritance principle on the momenta
and polarizations, i.e. we start form amplitudes in Minkowski
expressed with momenta and polarizations and then we implement on them
the projection to the orbifold.
In particular it is worth stressing that as there is one Killing
vector acting on the spacetime coordinates 
there is only
one common Killing vector action on all the momenta and polarizations
of each field
as discussed in the spin-1 and spin-2 cases.
Moreover this approach gives the complete answer only  tree level
amplitudes since inside the loops twisted states may be created in pairs.

The final result is that the open string amplitude with two tachyons
and the first massive (level 2) state diverges and there is no obvious
way of curing it since the divergence is also present in the Abelian sector.

The open string expansion we use is
\begin{equation}
  X(u,\bu)
  =\xzm -i\, 2\alpha'\, p\, \ln(|u|)
  +i \sqha \sum_{n\ne0} \frac{\alpha_n}{n} \left( u^{-n} + \bu^{-n}  \right)
.
\end{equation}

\subsection{Level 2 Massive State}
Before computing the amplitude we would like to review the possible
polarizations of the first massive state in open string.
The first massive vertex is
\begin{equation}
  V_M(x; k, S, \xi)=~
  :\!
  \left(
    \frac{i}{\sqda}
    \xi \cdot \partial^2_x X(x,x)
    +  \left(\frac{i}{\sqda}\right)^2
    S_{\mu\nu} \partial_x X^\mu(x,x) \partial_x X^\nu(x,x)
    \right)
    e^{i k \cdot X(x,x)}
    \!:~
,
\end{equation}
and the corresponding state is
\begin{align}
  \lim_{x\rightarrow 0}   V_M(x; k, S, \xi) |0 \rangle
  &=
    |k, S, \xi \rangle
    =
    \left(
    \xi \cdot \alpha_{-2}
    +
    \alpha_{-1} \cdot S \cdot \alpha_{-1}
    \right)
    | k\rangle
    .
  \end{align}
  The physical conditions read
  \begin{align}
    (L_0-1)    |k, S, \xi \rangle=0
    &\Rightarrow & \ap k^2=-1
                   \nonumber\\
    L_1    |k, S, \xi \rangle=0
    &\Rightarrow & S\cdot k+ \xi=0
                   \nonumber\\
    L_2    |k, S, \xi \rangle=0
    &\Rightarrow & k\cdot \xi+ \tr(S)=0
                   .
  \end{align}

  String gauge invariance allows to add
\begin{equation}
  L_{-1}( \chi \cdot \alpha_{-1} | k \rangle )
  =
  (
  \chi \cdot \alpha_{-2}
  +
  \chi \cdot \alpha_{-1}\,   k \cdot \alpha_{-1}
  )
  | k \rangle 
  ,
\end{equation}
subject to the physical constraints, i.e.
\begin{equation}
  \ap k^2=-1,
  ~~~~
  \chi \cdot k = 0
  .
\end{equation}
  Actually in critical string theory there is another gauge invariance
  generated by $L_{-2}+\frac{3}{2}L_{-1}^2$, in this case we can add
  a multiple of
  \begin{equation}
(    L_{-2}+\frac{3}{2}L_{-1}^2 ) | k \rangle 
  =
  (
  \frac{5}{2} k \cdot \alpha_{-2}
  +
  \frac{3}{2}  (  k \cdot \alpha_{-1} )^2
  +
  \frac{1}{2}  \alpha_{-1}^2
  )
  | k \rangle
  ,
\end{equation}
to set $a=0$.
Therefore the only non trivial d.o.f. are $S^{TT}$, i.e.
\begin{equation}
  \tr(S^{TT}) = k \cdot S^{TT} = \xi =0.
\end{equation}

  In view of the computation for the orbifold we can check that given
  $k=(\kp, \km, \kk, \vec k)$ ($-2\kp \km + \kk^2 +\vk^2=-1$)
  we can find a non trivial $S^{TT}$ with
  non vanishing components in the directions $\pm, 2$ only.
  We find in fact a two parameters family of solutions.
  The parameters may be taken to be $\Ss ++$ and $\Ss + 2$.
  Explicitly
  \begin{equation}
  \left(\begin{array}{c}
          \Ss + + \\
          \Ss + - \\
          \Ss + 2 \\
          \Ss - - \\
          \Ss - 2 \\
          \Ss 2 2
        \end{array}\right)
      =
      \left(\begin{array}{c}
              1 \\
              -\frac{\km}{\kp} \\
              0 \\
              \frac{\km (\km \kp -2 \kk^2) } {\kp^3} \\
              -2 \frac{\km \kk }{\kp^2} \\
              -2 \frac{\km}{\kp}
            \end{array}\right)
          \Ss ++
      +
        \left(\begin{array}{c}
                0 \\
                \frac{\kk}{\kp} \\
                1\\
              \frac{2 \kk (-\km \kp + \kk^2) } {\kp^3} \\
                \frac{\km \kp -2 \kk^2 } {\kp^2} \\
                2 \frac{\kk}{\kp}
        \end{array}\right)
      \Ss + 2
\end{equation}
  There is even a non trivial solution for the more special case
  $k=(\kp, \km= 1/\kp, \kk=0, \vec 0)$.

  Similarly using the expressions for $S^{ T T}$ in orbifold coordinates
  we check that there are two possible indepdendent polarizations
  $\cS_{v v}$ and $\cS_{v z}$ which correspond to the the ones used
  above.
  Then the non trivial solution is
  \begin{equation}
    \begin{pmatrix}
      \cS_{v v} \\
      \cS_{u v} \\
      \cS_{v z} \\
      \cS_{u u} \\
      \cS_{u z} \\
      \cS_{z z}       
    \end{pmatrix}
    =
    \begin{pmatrix}
      1 \\
      - \frac{r+ \vk^2}{2 \kp^2} \\
      0 \\
      \left( \frac{r+ \vk^2}{2 \kp^2} \right)^2 \\
      0 \\
      -2 \frac{r+ \vk^2}{2 \kp^2}
    \end{pmatrix}
    \cS_{v v}
    +
    \begin{pmatrix}
      0 \\
      -\frac{r+ \vk^2}{2 \kp^2} \\
      1 \\
      0 \\ 0\\ 0\\
    \end{pmatrix}
    \cS_{v z}
    .
  \end{equation}

  \subsection{Two Tachyons First Massive State Amplitude}
This Minkowskian
full amplitude is given by the sum of two color ordered ones as
\begin{equation}
  \cA_{TTM}=
  A_{T_{(1)} T_{(2)} M_{(3)}}\, \tr(T_{(1)} T_{(2)} T_{(3)})
  + A_{T_{(2)} T_{(1)} M_{(3)}}\, \tr(T_{(2)} T_{(1)} T_{(3)})
   ,
\end{equation}
where an easy computation gives
\begin{align}
  A_{T_{(1)} T_{(2)} M_{3)}}
  &=
    \langle\langle \kN1 |\,
    V_T(1; \kN2)\,
    (\alpha_{-1} \cdot S_{(3)}^{TT} \cdot \alpha_{-1}  |\kN3 \rangle)
    \nonumber\\
  &=
    \langle\langle \kN1 |\,
    e^{i\, \kN2 \cdot \xzm} e^{-\sqda \kN2 \cdot \alpha_{1}}
    (\alpha_{-1} \cdot S_{(3)}^{TT} \cdot \alpha_{-1}  |\kN3 \rangle)
    \nonumber\\
  &=
    (2\pi)^D \delta^D\left(\sum \kN i \right)~~
    (\sqda)^2\, \kN 2 \cdot S_{(3)}^{TT} \cdot \kN 2
    .
\end{align}
Because of transversality of $ S_{(3)}^{TT}$
the other term gives the same result of this
one, hence the final Minkowskian amplitude is
\begin{equation}
  \cA_{TTM}
  =
    (2\pi)^D \delta^D\left( \sum \kN i \right)~~
    2 (\sqda)^2\, \kN 2 \cdot S_{(3)}^{TT} \cdot \kN 2\,
    \tr\left( \{T_{(1)}, T_{(2)}\} T_{(3)}\right)
    .
\end{equation}
Then we can compute the orbifold amplitude as
\begin{multline}
  \cA_{TTM}
  =
  (2\pi)^{D-2}
  \delta^{D-3}\left( \sum \vkN i \right) \delta\left( \sum \kpN i \right)
  \nonumber\\
  2 (\sqda)^2\,
    \sum_{ \{\mN1, \mN2, \mN3\}\in \Z^3}\,
    \delta_{\mN3,1}\,
  (\Ki{\mN2}\kN 2) \cdot S_{(3)}^{TT} \cdot (\Ki{\mN2}\kN 2)
  \nonumber\\
    \delta\left( \sum (\Ki{\mN i}\kkN i \right)
    \delta\left( \sum (\Ki{\mN i}\kmN i \right)
    \,
    \tr\left( \{T_{(1)}, T_{(2)}\} T_{(3)}\right)
    .
\end{multline}
The previous amplitude can then be expressed using an overlap as
\begin{align}
  \cA_{TTM}
  &=
    2 (-i \sqda )^2\,
    &&
    \int_\Omega \dd^3x\,
      g^{\mu\nu}\,g^{\rho\sigma}\,
      \Psi^{[2]}_{[\kN{3}, \SsN{3}]\, \mu\rho}([x])\,
      \partial^2_{\nu\sigma} \Psi_{ [\kN{2}] }([x])\,
      \Psi_{ [\kN{1}] }([x])
     \nonumber\\
  &   &&   \tr\left( \{T_{(1)}, T_{(2)}\} T_{(3)}\right)
,   \nonumber\\
  &=
    2 (-i \sqda )^2\,
    &&
    \int_\Omega \dd^3x\,
      g^{\alpha\beta}\,g^{\gamma\delta}\,
      \Psi^{[2]}_{[\kN{3}, \SsN{3}]\, \alpha\gamma}([x])\,
      D_\beta \partial_{\delta} \Psi_{ [\kN{2}] }([x])\,
    \Psi_{ [\kN{1}] }([x])
    \nonumber\\
  &   &&   \tr\left( \{T_{(1)}, T_{(2)}\} T_{(3)}\right)
.
\end{align}
As discussed in the Section~\ref{sec:overlap} the last integral is divergent when
$\Ss++ =\cS_{v v}\ne 0$
and the divergence cannot be avoided even introducing a Wilson line
around $z$ since the amplitude involves an anticommutator which does
not vanish in the Abelian sector.

  \section{Scalar QED on \GNBO and Divergences}
\label{sect:genNOscalarQED}

As seen in the previous sections, the issues related to the vanishing volume of the compact directions lead to incurable divergences. We introduce the \GNBO by inserting one additional non compact direction with respect to the \NBO and show that divergences no longer occur.

As a parallel discussion to the \NBO, we introduce the geometry of the \GNBO and study scalar and spin-1 eigenfunctions to build the sQED on the orbifold. We then show how the presence of a non compact direction (we will stress the key differences from the \NBO\!\!) can cure the theory when considering amplitudes and overlaps.

\subsection{Geometric Preliminaries}
Consider Minkowski spacetime $\R^{1,D-1}$ and the change of coordinates from the lightcone set $( x^{\mu} ) = ( \xp, \xm, \xx, \xxx, \vex )$ to $( x^{\alpha} ) = ( u, v, w, z, \vex )$:
\begin{equation}
    \begin{split}
        &\begin{cases}
            \xm & = u
            \\
            \xp & = v + \frac{\Dx^2}{2} u ( z + w )^2 + \frac{\Dxx^2}{2} u ( z - w )^2
            \\
            \xx & = \Dx u ( z + w )
            \\
            \xxx & = \Dxx u ( z - w )
        \end{cases}
        \\
        \Leftrightarrow
        &\begin{cases}
            u & = \xm
            \\
            v & = \xp - \frac{1}{2 \xm} \left( (\xx)^2 + (\xxx)^2 \right)
            \\
            w & = \frac{1}{2\xm} \left( \frac{\xx}{\Dx} - 
                                        \frac{\xxx}{\Dxx} \right)
            \\
            z & = \frac{1}{2\xm} \left( \frac{\xx}{\Dx} +
                                        \frac{\xxx}{\Dxx} \right)
        \end{cases}
    \end{split}
    \label{eq:orbifold_coordinates}
\end{equation}
where we do not perform any change on the transverse coordinates $\vex$. The metric in these coordinates is non diagonal:
\begin{equation}
    \dd{s}^2 = - 2 \dd{u}\dd{v} 
             + ( \Dx^2 + \Dxx^2 ) u^2 ( \dd{w}^2 + \dd{z}^2 )
             + 2 ( \Dx^2 - \Dxx^2 ) u^2 \dd{w}\dd{z}
             + \eta_{ij}\dd{x}^i\dd{x}^j,
    \label{eq:orbifold_metric}
\end{equation}
and its determinant is:
\begin{equation}
    - \det{g} = 4 \Dx^2 \Dxx^2 u^4.
\end{equation}
From the previous expressions we can also derive the non vanishing Christoffel symbols:
\begin{equation}
    \begin{split}
        \Gamma^v_{ww} = \Gamma^v_{zz} & = ( \Dx^2 + \Dxx^2 ) u,
        \\
        \Gamma^v_{wz}                 & = ( \Dx^2 - \Dxx^2 ) u,
        \\
        \Gamma^w_{uw} = \Gamma^z_{uz} & = \frac{1}{u},
    \end{split}
\end{equation}
which however produce a vanishing Ricci tensor and curvature scalar, since we are considering Minkowski spacetime anyway.

We now introduce the \GNBO by identifying points in space along the orbits of the Killing vector:
\begin{equation}
    \begin{split}
        \kappa & = - 2 \pi i ( \Dx J_{+2} + \Dxx J_{+3} )
        \\
               & = 2 \pi ( \Dx x^2 + \Dxx x^3 ) \partial_+
                 + 2 \pi \Dx x^- \partial_2 
                 + 2 \pi \Dxx x^- \partial_3
        \\
               & = 2 \pi \dz
    \end{split}
\end{equation}
in such a way that
\begin{equation}
    x^{\mu} \sim e^{n\kappa} x^{\mu}, \qquad n \in \Z
\end{equation}
leads to the identifications
\begin{align}
x=
        \begin{pmatrix}
        \xm
         \\
        \xx
        \\
        \xxx
        \\
        \xp
        \\
        \vex
\end{pmatrix}
\equiv
\Ki{n} x=
 \begin{pmatrix}
        \xm
        \\
        \xx + 2 \pi n \Dx\xm
        \\
        \xxx + 2 \pi n \Dxx\xm
         \\
        \xp + 2 \pi n \Dx\xx
            + 2 \pi n \Dxx\xxx
            + (2 \pi n)^2 \frac{\Dx^2+\Dxx^2}{2} \xm
         \\
        \vex
\end{pmatrix},
\end{align}
or to the simpler
\begin{equation}
    ( u, v, w, z ) \sim ( u, v, w, z + 2 \pi n )
\label{eq:orbifold_identifications}
\end{equation}
using the map to the orbifold coordinates \eqref{eq:orbifold_coordinates} where the Killing vector $\kappa = 2 \pi \dz$ does not depend on the local spacetime configuration.
As in the previous case, the difference between Minkowski spacetime and the \GNBO is therefore global.

The geodesic distance between the n-th copy and the base point on the orbifold can be computed in any set of coordinates and is:
\begin{equation}
    \Delta s^2_{(n)} = ( \Dx^2 + \Dxx^2 ) ( 2 \pi n \xm )^2 \ge 0.
\end{equation}
Closed time-like curves are therefore avoided on the \GNBO\!\!, but there are closed null curves on the surface $\xm = u = 0$ where the Killing vector $\kappa$ vanishes.

\subsection{Free Scalar Field}

In order to build a quantum theory on the \GNBO using Feynman's approach to quantization, we first solve the eigenvalue equations for the fields and then build their off-shell expansion. We start from a complex scalar field and then consider the free photon before moving to the sQED interactions on the \GNBO\!\!.

Consider the action for a complex scalar field:
\begin{equation}
    \begin{split}
        \act{\text{scalar kin}}
                & = \int_{\Omega} \dd^D x \sqrt{-\det{g}}
                     \left( -g^{\mu\nu} \partial_{\mu} \phi^* \partial_{\nu} \phi 
                            - M^2 \phi^* \phi \right)
        \\
                & =
                \int_{\R^{D-4}} \dd^{D-4} \vex
                \int_{-\infty}^{+\infty} \dd u
                \int_{-\infty}^{+\infty} \dd v
                \int_{-\infty}^{+\infty} \dd w
                \int_0^{2\pi} \dd z
                ~~
                2 \Abs{\Dx \Dxx} u^2
        \\
                & \times
                \Bigg[ \du \phi^* \dv \phi + \dv \phi^* \du \phi
                - \frac{1}{4 u^2}
                  \Bigg(
                    \Big(
                        \frac{1}{\Dx^2} + \frac{1}{\Dxx^2}
                    \Big)
                    \Big(
                        \dw \phi^* \dw \phi
                        +
                        \dz \phi^* \dz \phi
                    \Big)
        \\
                & +
                    \Big(
                        \frac{1}{\Dx^2} - \frac{1}{\Dxx^2}
                    \Big)
                    \Big(
                        \dw \phi^* \dz \phi
                        +
                        \dz \phi^* \dw \phi
                    \Big)
                  \Bigg)
                  -
                  \eta^{ij} \partial_i \phi^* \partial_j \phi
                  -
                  M^2 \phi^* \phi
               \Bigg].
    \end{split}
\end{equation}
As in the case of the \NBO\!\!, the solutions to the equations of motion are necessary to provide the modes of the quantum fields. We study the eigenvalue equation $\square \phi_r = r \phi_r$, where $r$ is $2 k_+ k_- - \vk$ by comparison with the flat case ($k$ is the momentum associated to the flat coordinates). We therefore need solve:
\begin{equation}
    \begin{split}
        \Bigg\lbrace
            &
            -2 \du \dv - \frac{2}{u} \dv
            + \frac{1}{4 u^2}
            \Big[
                \left( \frac{1}{\Dx^2} + \frac{1}{\Dxx^2} \right)
                \left( \dw^2 + \dz^2 \right)
        \\
            &
                +
                2
                \left( \frac{1}{\Dx^2} - \frac{1}{\Dxx^2} \right)
                \dw \dz          
            \Big]
            +
            \eta^{ij} \partial_i \partial_j
            - r
        \Bigg\rbrace \phi_r
        = 0.
    \end{split}
    \label{eq:scalar_eom}
\end{equation}
To this purpose, we introduce a Fourier transformation over $v, w, z, \vex$:
\begin{equation}
    \phi_r( u, v, w, z, \vex)
    =
    \sum\limits_{l \in \Z}
    \int_{\R^{D-4}} \dd^{D-4} \vk
    \int_{-\infty}^{+\infty} \dd k_+
    \int_{-\infty}^{+\infty} \dd p
    ~~
    e^{i\left( \kp v + \kw w + \mm z + \vk \cdot \vex \right)}
    \tpkmkrgen(u),
\end{equation}
where we defined $\kp, \kw, \mm, \vk$ as associated momenta to $v, w, z, \vex$ respectively, and we find:
\begin{equation}
    \phi_{\kmkrgen}( u, v, w, z, \vex )
    =
    e^{i\left( \kp v + \kw w + \mm z + \vk \cdot \vex \right)}
    \tpkmkrgen( u ).
\end{equation}
where
\begin{equation}
    \tpkmkrgen( u )
    =
    \frac{1}{2 \sqrt{(2 \pi)^D \abs{\Dx \Dxx \kp}}}~
    \frac{1}{\Abs{u}}
    e^{-i\left( \frac{1}{8 \kp u}
                \left[ \frac{(l + p)^2}{\Dx^2} + \frac{(l - p)^2}{\Dxx^2} \right]
                -
                \frac{\vk^2 + r}{2 \kp} u
         \right)}.
\end{equation}
These solutions present the right normalization, as we can verify through the product:
\begin{equation}
    \begin{split}
        &
        \left( \phi_{\kmkrgenN{1}},\, \phi_{\kmkrgenN{2}} \right)
        =
        2 \Abs{\Dx \Dxx}
        \\
        & \times
        \int_{\R^{D-4}} \dd^{D-4} \vex
        \int_{\R^3} \dd u~ \dd v~ \dd w~
        \int_0^{2\pi} \dd z~
        u^2~
        \phi_{\kmkrgenN{1}}~
        \phi_{\kmkrgenN{2}}
        \\
        & = \delta^{D - 4}( \vkN{1} + \vkN{2} )~
            \delta( \kpN{1} + \kpN{2} )~
            \delta( \kwN{1} + \kwN{2} )~
            \delta( \rN{1} + \rN{2} )~
            \delta_{\mmN{1}, \mmN{2}}.
    \end{split}
\end{equation}

Then we have the off-shell expansion:
\begin{equation}
    \begin{split}
        \phi_r( u, v, w, z, \vex )
        & =
        \frac{1}{2 \sqrt{( 2 \pi )^D \Abs{\Dx \Dxx \kp}}}
        \sum\limits_{l \in \Z}
        \int_{\R^{D-4}} \dd^{D-4} \vk
        \int_{-\infty}^{+\infty} \dd k_+
        \int_{-\infty}^{+\infty} \dd p
        \int_{-\infty}^{+\infty} \dd r
        \\
        & \times
        \frac{\mathcal{A}_{\kmkrgen}}{\Abs{u}}
        e^{i\left(
                \kp v + \kw w + \mm z + \vk \cdot \vex
                - \frac{1}{8 \kp u}
                  \left[
                    \frac{(l + p)^2}{\Dx^2} + \frac{(l - p)^2}{\Dxx^2}
                  \right]
                + \frac{\vk^2 + r}{2 \kp} u
          \right)}.
    \end{split}
\end{equation}

\subsection{Free Photon Action}

We then study the action of the free photon field $a$ using the Lorenz gauge which in the orbifold coordinates it reads:
\begin{equation}
    \begin{split}
        D^\alpha a_\alpha
        &
        =
        - \frac{2}{u} a_{v}
        - \dv a_u
        - \du a_v
    \\
        &
        + \frac{1}{4 u^2}
        \bigg(
          \Big(
              \frac{1}{\Dx^2} + \frac{1}{\Dxx^2}
          \Big)
          \Big(
              \dw a_w + \dz a_z
          \Big)
          +
          \Big(
              \frac{1}{\Dx^2} - \frac{1}{\Dxx^2}
          \Big)
          \Big(
              \dw a_z + \dz a_w
          \Big)
        \bigg)
    \\
        &
        + \eta^{ij} \partial_i a_j
        = 0.
    \end{split}
\end{equation}

We then solve the eigenvalue equations
$\left( \square a_r \right)_{\nu} = r~ a_{r\,\nu}$,
which in components read:
\begin{equation}
    \begin{split}
        \left( \square a_r \right)_u
        & =
        \frac{2}{u^2} a_{r\,v}
        - \frac{1}{2 u^3}
        \left[
            \left( \frac{1}{\Dx^2} + \frac{1}{\Dxx^2} \right)
            \left( \dw a_{r\,w} + \dz a_{r\,z} \right)
            +
            \left( \frac{1}{\Dx^2} - \frac{1}{\Dxx^2} \right)
            \left( \dw a_{r\,z} + \dz a_{r\,w} \right)
        \right]
        \\
        & +
        \left\lbrace
            - 2 \du \dv
            - \frac{2}{u} \dv
            + \frac{1}{4 u^2}
            \left[
                \left( \frac{1}{\Dx^2} + \frac{1}{\Dxx^2} \right)
                \left( \dw^2 + \dz^2 \right)
                +
                \left( \frac{1}{\Dx^2} - \frac{1}{\Dxx^2} \right)
                2 \dw \dz
            \right]
            + \nabla^2_T
        \right\rbrace
        a_{r\,u}
        ,
        \\
        \left( \square a_r \right)_v
        & =
        \left\lbrace
            - 2 \du \dv
            - \frac{2}{u} \dv
            + \frac{1}{4 u^2}
                \left[
                    \left( \frac{1}{\Dx^2} + \frac{1}{\Dxx^2} \right)
                    \left( \dw^2 + \dz^2 \right)
                    +
                    \left( \frac{1}{\Dx^2} - \frac{1}{\Dxx^2} \right)
                    2 \dw \dz
                \right]
            + \nabla^2_T
        \right\rbrace
        a_{r\,v}
        ,
        \\
        \left( \square a_r \right)_w
        & =
        - \frac{2}{u} \dw a_{r\,v}
        \\
        & +
        \left\lbrace
            - 2 \du \dv 
            + \frac{1}{4 u^2}
            \left[
                \left( \frac{1}{\Dx^2} + \frac{1}{\Dxx^2} \right)
                \left( \dw^2 + \dz^2 \right)
                + 
                \left( \frac{1}{\Dx^2} - \frac{1}{\Dxx^2} \right)
                2 \dw \dz
            \right]
            + \nabla^2_T
        \right\rbrace
        a_{r\,w}
        ,
        \\
        \left( \square a \right)_z
        & =
        - \frac{2}{u} \dz a_{r\,v}
        \\
        & +
        \left\lbrace
            - 2 \du \dv
            + \frac{1}{4 u^2}
            \left[
                \left( \frac{1}{\Dx^2} + \frac{1}{\Dxx^2} \right)
                \left( \dw^2 + \dz^2 \right)
                + 
                \left( \frac{1}{\Dx^2} - \frac{1}{\Dxx^2} \right)
                2 \dw \dz
            \right]
            + \nabla^2_T
        \right\rbrace
        a_{r\,z}
        ,
        \\
        \left( \square a \right)_i
        & =
        \left\lbrace
            - 2 \du \dv
            - \frac{2}{u} \dv
            + \frac{1}{4 u^2}
                \left[
                    \left( \frac{1}{\Dx^2} + \frac{1}{\Dxx^2} \right)
                    \left( \dw^2 + \dz^2 \right)
                    +
                    \left( \frac{1}{\Dx^2} - \frac{1}{\Dxx^2} \right)
                    2 \dw \dz
                \right]
            + \nabla^2_T
        \right\rbrace
        a_{r\,i}
        ,
    \end{split}
\end{equation}
where $\nabla^2_T = \eta^{ij} \partial_i \partial_j$ is the Laplace operator in the transverse coordinates $\vex$. These equations can be solved using standard techniques through a Fourier transform:
\begin{equation}
    a_{r\, \alpha}( u, v, w, z, \vex)
    =
    \sum\limits_{l \in \Z}
    \int_{\R^{D-4}} \dd^{D-4} \vk
    \int_{-\infty}^{+\infty} \dd k_+
    \int_{-\infty}^{+\infty} \dd p
    ~~
    e^{i\left( \kp v + \kw w + \mm z + \vk \cdot \vex \right)}
    \takmkrgen{\alpha}(u).
\end{equation}
We first solve the equations for $\takmkrgen{v}$ and $\takmkrgen{i}$ since they are identical to the scalar equation \eqref{eq:scalar_eom}. We then insert their solutions as sources for the equations for $\takmkrgen{u}$, $\takmkrgen{w}$ and $\takmkrgen{z}$. The solutions can be written as the expansion:
\begin{equation}
    \begin{split}
        \parallel \tilde{a}_{\kmkrgen\, \alpha}(u) \parallel
        =
        \begin{pmatrix}
            \tilde{a}_u
            \\
            \tilde{a}_v
            \\
            \tilde{a}_w
            \\
            \tilde{a}_z
            \\
            \tilde{a}_i
        \end{pmatrix}
        & =
        \sum_{\underline{\alpha}
              \in 
              \{ \underline{u}, 
                 \underline{v}, 
                 \underline{w},
                 \underline{z},
                 \underline{i} 
              \}
             }
        \cNEgen{\alpha}\,
        \parallel \tilde{a}^{\underline{\alpha}}_{\kmkrgen\, \alpha}(u) \parallel
        \\
        & =
        \cNEgen{u}\,
        \begin{pmatrix}
            1
            \\
            0
            \\
            0
            \\
            0
            \\
            0
        \end{pmatrix}\,
        \tpkmkrgen
        \\
        & +
        \cNEgen{v}\,
        \begin{pmatrix}
            \frac{i}{2 \kp u}
            +
            \frac{1}{8 \kp^2 u^2}
            \left( \frac{(l + p)^2}{\Dx^2} + \frac{(l - p)^2}{\Dxx^2} \right)
            \\
            1
            \\
            \frac{p}{\kp}
            \\
            \frac{l}{\kp}
            \\
            0
        \end{pmatrix}\,
        \tpkmkrgen
        \\
        & +
        \cNEgen{w}\,
        \begin{pmatrix}
            \frac{1}{4 \kp \abs{u}}
            \left( \frac{l + p}{\Dx^2} - \frac{l - p}{\Dxx^2} \right)
            \\
            0
            \\
            \abs{u}
            \\
            0
            \\
            0
        \end{pmatrix}\,
        \tpkmkrgen
        \\
        & +
        \cNEgen{z}\,
        \begin{pmatrix}
            \frac{1}{4 \kp \abs{u}}
            \left( \frac{l + p}{\Dx^2} + \frac{l - p}{\Dxx^2} \right)
            \\
            0
            \\
            0
            \\
            \abs{u}
            \\
            0
        \end{pmatrix}\,
        \tpkmkrgen
        \\
        & +
        \cNEgen{j}\,
        \begin{pmatrix}
            0
            \\
            0
            \\
            0
            \\
            0
            \\
            \delta_{\underline{i j}}
        \end{pmatrix}\,
        \tpkmkrgen
    \end{split}
\end{equation}
Consider the Fourier transformed functions:
\begin{equation}
    a^{\underline{\alpha}}_{\kmkrgen\, \alpha}( u, v, w, z, \vex )
    =
    e^{i(\kp v + \kw w + \mm z + \vk \cdot \vex)}
    \tilde{a}^{\underline{\alpha}}_{\kmkrgen\, \alpha}( u ),
\end{equation}
then we can expand the off shell fields as
\begin{equation}
    a_{\alpha}(x)
    =
    \sum_{\mm \in \Z}\,
    \int \dd^{D-4} \vk\,
    \int \dd \kp\,
    \int \dd p\,
    \int \dd r\,
    \sum_{\underline{\alpha}
          \in
          \{ \underline{u},
             \underline{v},
             \underline{w},
             \underline{z},
             \underline{i}
          \}
         }
    \cNE{\alpha}\,
    {a}^{\underline{\alpha}}_{\kmkrgen\, \alpha}(x).
\end{equation}

We can compute the normalization as:
\begin{equation}
    \begin{split}
        \left( a_{(1)},\, a_{(2)} \right)
        & =
        \int_{\R^{D-4}} \dd^{D-4} \vex
        \int_{\R^3} \dd u~ \dd v~ \dd w
        \int_0^{2\pi} \dd z~
        2 \abs{\Dx \Dxx} u^2~
        \\
        & \times
        \left(
            g^{\alpha\beta}~
            a_{\kmkrgenN{1}\, \alpha}~
            a_{\kmkrgenN{2}\, \beta}
        \right)
        \\
        & =
        \delta^{D-4}( \vkN{1} + \vkN{2} )~
        \delta( p_{(1)} + p_{(2)} )~
        \delta( \kpN{1} + \kpN{2} )~
        \delta_{l_{(1)} + l_{(2)}, 0}
        \delta( r_1 - r_2 )
        \\
        & \times
        \cE_{\kmkrgenN{1}} \circ \cE_{\kmkrgenN{2}},
    \end{split}
\end{equation}
where
\begin{equation}
    \begin{split}
        \cE_{(1)} \circ \cE_{(2)}
        & =
        -\cE_{(1)\, u}~\cE_{(2)\, v}
        -\cE_{(1)\, v}~\cE_{(2)\, u}
        \\
        &
        +\frac{1}{4}
        \Bigg[
            \bigg( \frac{1}{\Dx^2} + \frac{1}{\Dxx^2} \bigg)
            \bigg( \cE_{(1)\, w}~\cE_{(2)\, w}
                   +
                   \cE_{(1)\, z}~\cE_{(2)\, z}
            \bigg)
        \\
            & +
            \bigg( \frac{1}{\Dx^2} - \frac{1}{\Dxx^2} \bigg)
            \bigg( \cE_{(1)\, w}~\cE_{(2)\, z}
                   +
                   \cE_{(1)\, z}~\cE_{(2)\, w}
            \bigg)
        \Bigg]
    \end{split}    
\end{equation}
is independent of the coordinates. The Lorenz gauge now reads:
\begin{equation}
    \eta^{ij} k_i \, \cE_{{\kmkrgen}j}
    -
    \kp
    \cNEgen{u}
    -
    \frac{\vk^2 + r}{2 \kp}
    \cNEgen{v}
    = 0.
\end{equation}
As in the previous case, it does not pose any constraint on the transverse polarizations $\cNEgen{w}$ and $\cNEgen{z}$.

\subsection{Cubic Interaction}

As previously studied on the \NBO\!\!, we can now show the sQED 3-points vertex computation using the previously computed eigenmodes. The presence of a continuous momentum in the non compact direction plays a major role in saving the convergence of the integrals. In the case of the \GNBO we find:
\begin{equation}
    \begin{split}
        \act{\text{cubic}}     
        & =
        \int_{\Omega} \dd^D x \sqrt{-\det{g}}
        \left( - i e g^{\mu\nu} a_{\mu}
               \left( \phi^* \partial_{\nu} \phi
                      -
                      \partial_{\nu} \phi^* \phi
               \right)
        \right)
        \\
        & =
        \prod\limits_{i = 1}^3
        \sum\limits_{l_{(i)} \in \Z}~
        \int \dd^{D-4} \vkN{i}
        \int \dd \kpN{i}
        \int \dd p_{(i)}
        \int \dd r_{(i)}
        \\
        & \times
        ( 2 \pi )^{D-1}~
        \delta^{D-4}( \sum\limits_{i=1}^3 \vkN{i} )~
        \delta( \sum\limits_{i=1}^3 p_{(i)} )~
        \delta( \sum\limits_{i=1}^3 \kpN{i} )~
        \delta_{\sum\limits_{i=1}^3 l_{(i)},\, 0}
        \\
        & \times
        e
        \mathcal{A}^*_{\MINUSkmkrgenN{2}}
        \mathcal{A}_{\kmkrgenN{3}}
        \\
        & \times
        \Bigg\lbrace
            \cE_{\kmkrgenN{1}\, u}~
            \kpN{2}~
            \INT{3}{0}
            \\
            & +
            \cE_{\kmkrgenN{1}\, v}~
            \Bigg[
                \left( \frac{\vkN{2}^2 + r_{(2)}}{2 \kpN{2}} \right)~
                \INT{3}{0}
                +
                i \frac{\kpN{2}}{\kpN{1}}~
                \INT{3}{-1}
                \\
                & +
                \frac{\kpN{2}}{8}
                \left[
                    \frac{1}{\Dx^2}
                    \left( 
                        \frac{l_{(1)} + p_{(1)}}{\kpN{1}}
                        +
                        \frac{l_{(2)} + p_{(2)}}{\kpN{2}}
                    \right)^2
                    +
                    \frac{1}{\Dxx^2}
                    \left( 
                        \frac{l_{(1)} - p_{(1)}}{\kpN{1}}
                        +
                        \frac{l_{(2)} - p_{(2)}}{\kpN{2}}
                    \right)^2
                \right]~
                \INT{3}{-2}
            \Bigg]
            \\
            & +
            \left( \cE_{\kmkrgenN{1}\, w} - \cE_{\kmkrgenN{1}\, z} \right)
            \Bigg[
                \frac{1}{\Dx^2}
                \left( 
                    \frac{\kpN{1} ( l_{(2)} + p_{(2)} )
                          +
                          \kpN{2} ( l_{(1)} + p_{(1)} )} {\kpN{1}}
                \right)
                \\
                & -
                \frac{1}{\Dxx^2}
                \left( 
                    \frac{\kpN{1} ( l_{(2)} - p_{(2)} )
                          +
                          \kpN{2} ( l_{(1)} - p_{(1)} )} {\kpN{1}}
                \right)
            \Bigg]~
            \MINT{3}{-1}
            \\
            & +
            \Bigl( (2) \leftrightarrow (3) \Bigr)
        \Bigg\rbrace
    \end{split}
\end{equation}
where we defined:
\begin{equation}
    \begin{split}
        \INT{N}{\nu}
        & =
        \int_{\R} \dd u~ 2 \abs{\Dx \Dxx} u^2~ u^{\nu}~ \prod\limits_{i=1}^N \tpkmkrgenN{i},
        \\
        \MINT{N}{\nu}
        & =
        \int_{\R} \dd u~ 2 \abs{\Dx \Dxx} u^2~ \abs{u}^{\nu}~ \prod\limits_{i=1}^N \tpkmkrgenN{i}.
    \end{split}
\end{equation}

Differently fron the \NBO case, we do not need any particular regularization while treating these integrals, as the singular factors in the phase only take continuous values due to the presence of the $p$ continuous momentum. The phase does not have isolated zeros and the integral can be given a distributional interpretation thus curing possible divergences.

\subsection{Quartic Interactions}

As for the NBO, we consider the quartic interaction for the sQED action:
\begin{equation}
    \begin{split}
        \act{\text{quartic}}
        & =
        \int_{\Omega} \dd^D x \sqrt{-\det{g}}
        \left( e^2~ g^{\mu\nu}~ a_{\mu} a_{\nu} \abs{\phi}^2
               -
               \frac{g_4}{4} \abs{\phi}^4
        \right)
        \\
        & \times
        \prod\limits_{i = 1}^3
        \left( \frac{1}{4 \pi \sqrt{( (2\pi)^D \abs{\Dx \Dxx \kpN{i}}}} \right)
        \sum\limits_{l_{(i)} \in \Z}
        \int \dd^{D-4} \vkN{i}
        \int \dd \kpN{i}
        \int \dd p_{(i)}
        \int \dd r_{(i)}
        \\
        & \times
        ( 2 \pi )^{D-1}~
        \delta^{D-4}( \sum\limits_{i=1}^3 \vkN{i} )~
        \delta( \sum\limits_{i=1}^3 p_{(i)} )~
        \delta( \sum\limits_{i=1}^3 \kpN{i} )~
        \delta_{\sum\limits_{i=1}^3 l_{(i)},\, 0}
        \\
        & \times
        \Bigg\lbrace
            e^2
            \mathcal{A}^*_{\MINUSkmkrgenN{3}}
            \mathcal{A}_{\kmkrgenN{4}}
            \\
            & \times
            \Bigg[
                \cE_{\kmkrgenN{1}} \circ \cE_{\kmkrgenN{2}}~
                \INT{4}{0}
                \\
                & - i
                \cE_{\kmkrgenN{1}\, v}\,
                \cE_{\kmkrgenN{2}\, v}
                \Bigg(
                    \left( \frac{1}{\kpN{1}} + \frac{1}{\kpN{2}} \right)~
                    \INT{4}{-1}
                    \\
                    & - i
                    \left(
                        \frac{\mathcal{G}_{+\,(1,2)}}{\Dx^2}
                        +
                        \frac{\mathcal{G}_{-\,(1,2)}}{\Dxx^2}
                    \right)~
                    \INT{4}{-2}
                \Bigg)
                \\
                & +
                \frac{1}{4}
                \Bigg(
                    \tilde{\cE}_{+\,(1,2)}~
                    \frac{\mathcal{G}_{+\,(1,2)}}{\Dx^2}
                    -
                    \tilde{\cE}_{-\,(1,2)}~
                    \frac{\mathcal{G}_{-\,(1,2)}}{\Dx^2}
                \Bigg)~
                \MINT{4}{-1}
            \Bigg]
            \\
            & -
            \frac{g_4}{4}
            \mathcal{A}^*_{\MINUSkmkrgenN{1}}
            \mathcal{A}^*_{\MINUSkmkrgenN{2}}
            \\
            & \times
            \mathcal{A}_{\kmkrgenN{3}}
            \mathcal{A}_{\kmkrgenN{4}}
            \INT{4}{0}
        \Bigg\rbrace,
    \end{split}
\end{equation}
where we defined:
\begin{equation}
    \begin{split}
        \mathcal{G}_{\pm\,\left( a, b \right)}
        & =
        \frac{l_{(a)} \pm p_{(a)}}{\kpN{a}}
        -
        \frac{l_{(b)} \pm p_{(b)}}{\kpN{b}},
        \\
        \tilde{\cE}_{\pm\,\left( a, b \right)}
        & =
        \cE_{\kmkrgenN{a}\, v}
        \left( \cE_{\kmkrgenN{b}\, w} \pm \cE_{\kmkrgenN{b}\, z} \right)
        \\
        & -
        \cE_{\kmkrgenN{b}\, v}
        \left( \cE_{\kmkrgenN{a}\, w} \pm \cE_{\kmkrgenN{a}\, z} \right)
    \end{split}
\end{equation}
for simplicity.

Differently from the the case of the NBO, the theory on the GNBO is not ill defined as previously discussed. The lack of isolated zeros is key to the convergence of the integral which can be interpreted as distributions in this case.

\subsection{Resurgence of Divergences}

Looking back at the metric \eqref{eq:orbifold_metric} and at the identifications \eqref{eq:orbifold_identifications} where we compactified only the coordinate $z$ through the Killing vector $2 \pi \dz $ it seems reasonable to wonder what would happen if we acted in the same way over $w$, since $2 \pi \dw $ is a Killing vector as well and it commutes with $2\pi\dz$. 
However the lesson we learnt from our whole study on \NBO and \GNBO is that in absence of at least one continuous transverse direction it's not possible to avoid the divergences associated with discrete zero energy modes and this is exactly what happens.

\section{Quick Analysis of the BO}
\label{sect:BO}
In this section we would like to quickly show the analysis
performed in the previous sections for the \NBO but in the case of the \BO\!\!.
The results are not very different apart from the fact that
divergences are milder, in fact it is possible to construct the full
sQED but nevertheless it is impossible to consider higher derivative
terms in the effective theory and some three point amplitudes with a massive
state diverge.

%
%


\subsection{Geometric Preliminaries}
In $\R^{1,1}$ we consider the change of coordinates:
\begin{align}
  \begin{cases}
    \xp = t \LLP+
    \\
    \xm = \nsm\, t \LLP-
  \end{cases}
  \Leftrightarrow
  \begin{cases}
    t = \sgn(\xp)\, \sqrt{ |\xp \xm| }
    \\
    \vvphi = \frac{ 1 }{ 2 \loLd} \log \left| \frac{\xp}{\xm} \right|
    \\
    \nsm = \sgn( \xp \xm)
  \end{cases}
\end{align}
where $\nsm=\pm1$ and $t,~\vvphi \in \R$. The metric reads
\begin{align}
  \dd s^2 &= -2 \dd \xp\, \dd \xp
            \nonumber\\
  &= -2 \nsm ( \dd t ^2 - (\loLd t)^2\, \dd \vvphi^2 )
    ,
\end{align}
and its determinant is:
\begin{equation}
    -\det g = 4 \loLd^2 t^2.
\end{equation}
In orbifold coordinates the non vanishing Christoffel symbols are:
\begin{equation}
    \Gamma^t_{\vvphi\,\vvphi}= \loLd^2 t,
    \qquad
    \Gamma^\vvphi_{t\,\vvphi}= \frac{1}{t}.
\end{equation}

Using the orbifold coordinates $(t,~\vvphi)$, the \BO is obtained by requiring the identification $\vvphi\equiv \vvphi+2\pi$ along the orbit of the global Killing vector $\kappa_{\vvphi} = 2 \pi \partial_{\vvphi}$. We will therefore use the recurrent parameter $\Lambda=e^{2\pi \Delta}$ in what follows.

\subsection{Free Scalar Action}
  The action for a complex scalar $\phi$ is given by
  \begin{align}
    S_{\mbox{scalar kin}}
    =&
       \int d^D x\,
       \sqrt{- \det g}
      \Bigl(
      -g^{\mu\nu} \partial_\mu \phi^* \partial_\nu \phi
      -M^2  \phi^* \phi
      \Bigr)
      \nonumber\\
    =
     &
      \sum_{\nsm\in\{\pm 1\}}
      \int d^{D-2} \vec x\,
      \int_{-\infty}^{+\infty} d t\, \int_0^{2\pi} d \phi\,
      \loLd |t|
       \nonumber\\
      \Biggl(
       \oh \nsm\,
       \partial_t \phi^*\,\partial_t \phi\,       &
       +
       \oh \nsm  
      \frac{1}{ (\loLd t)^2}  \partial_\vvphi \phi^*\,\partial_\vvphi  \phi\,
      -
      \partial_i \phi^* \partial_i \phi
      -
      M^2 \phi^* \phi
      \Biggr)
      ,
    \end{align}
As before we solve the associated eigenfunction
problem for the d'Alembertian operator
    \begin{align}
      -\oh \nsm \partial_t^2 \phi_r
      -
      \oh \nsm
      \frac{1}{t} \partial_t \phi_r
      +
      \oh \nsm
      \frac{1}{ (\loLd\, t)^2} \partial_\vvphi^2 \phi_r
      +
      \partial_i^2 \phi_r
      =
      r \phi_r
      .
    \end{align}
    with
    \begin{equation}
      r= 2 \kp \km - \vk^2= 2 \vsm m^2  -\vk^2      
      \end{equation}
where
      for later convenience (see the transformation of $k$ under the
      induced action of the Killing vector \eqref{eq:BO_kpkp_equivalence})
    we parameterize the momenta as the coordinates
\begin{align}
  \begin{cases}
    \kp = m\, \LLB+
    \\
    \km = \vsm m\, \LLB-
  \end{cases}
\Leftrightarrow
  \begin{cases}
    m = \sgn(\kp)\, \sqrt{ |\kp \km| }
    \\
    \beta = \frac{ 1}{ 2 \loLd} \log \left| \frac{\kp}{\km} \right|
    \\
    \vsm = \sgn( \kp \km)
  \end{cases}
  \label{eq:kpkm_parametrization}
\end{align}
where $\vsm=\pm1$ and $m,~\beta\in \R$.
To solve the problem we use standard techniques and perform the Fourier
transform wrt $\vec x$ and $\phi$ as 
    \begin{equation}
      \phi(t, \vvphi, \vec x)
      = \int d^{D-2} \vec x\,
      \sum_{l\in \Z}
      e^{i \vk \cdot \vec x}
      e^{i l \vvphi} \HH(t)
        ,
      \end{equation}
so that the new function $\HH$ satisfies
    \begin{align}
       \partial_t^2 \HH
      +
      \frac{1}{t} \partial_t \HH
      +
      \left[
      \frac{l^2}{ (\loLd\, t)^2} 
      +
      2 \nsm
      (  r+ \vk^2)
      \right]
      \HH
      =
      0
      ,
    \end{align}
    which upon the introduction of the natural quantities (see also
    \eqref{eq:BO_PSI0_tau_lambda} for an explanation of the naturalness of $\lambda$)
    \begin{equation}
      \tau=m t,~~~~
      \lambda=e^{\Delta(\vvphi+\beta)},~~~~
      \hspm=\nsm \vsm,
      \end{equation}
    shows that the actual dependence on parameters is
    \begin{equation}
      \HH(t)= \tplsi(\tau)
      ,
      \end{equation}
      so that
    \begin{align}
       \partial_\tau^2 \tplsi
      +
      \frac{1}{\tau} \partial_\tau \tplsi
      +
      \left[
      \frac{l^2}{ (\loLd\, \tau)^2} 
      +
      4 \hspm
      \right]
      \tplsi
      =
      0
      .
      \label{eq:BO_eq_diff_tilde_phi}
    \end{align}
The  solutions have asymptotics
\begin{equation}
    \tplsi \sim
  \begin{cases}
    A_+ |\tau|^{i \frac{l}{\Delta}}
    +
    A_- |\tau|^{-i \frac{l}{\Delta}}
    &
    l \ne 0
    \\
    A_+ \log(|\tau|) + A_-
    & l=0
  \end{cases}
  ,
  \label{eq:BO_asymtotics}
  \end{equation}
  and we will be more concerned on the $l=0$ case as before.
  

    
\subsection{Eigenmodes on \BO from Covering Space}
We now repeat the essential part of the analysis
performed in the \NBO case.
As in the \NBO case we use the wording wave function and not the
eigenfunction because eigenfunctions for non scalar states require some 
constraints on polarizations which we do not impose.

\subsubsection{Spin 0}
We start as usual with the Minkowskian wave function and we write only
the dependence on $\xp$ and $\xm$ since all the other coordinates are spectators
\begin{align}
  \psi_{\BOkk}(\xp,\xm)
  &=
    e^{i\left( \kp \xp + \km \xm  \right)}
    \nonumber\\
  =\psi_{\BOkk}(t, \vvphi, \nsm)
  &
    =
    e^{i m t \left[ \LLPB+
    +
    \hspm t \LLPB-
    \right]}
    .
  \end{align}
We can compute the wave function on the orbifold by summing over all
images
\begin{align}
  \Psi_{[\BOkk]}([\xp,\xm])
  &=
  \sum_{n\in\Z}
    \psi_{\BOkk}( \Ki{n }(\xp,\xm))
    \nonumber\\
  &=
  \sum_{n\in\Z}
    \psi_{\BOkk}(\xp e^{2\pi\Delta\,n}
    , \xm e^{-2\pi\Delta\,n})
    \nonumber\\
  &=
  \sum_{n\in\Z}
    e^{i\left\{
    [ \kp e^{2\pi\Delta\,n} ] \xp
    +
   [\km e^{-2\pi\Delta\,n}] \xm
     \right\}}
    \nonumber\\
    &=
  \sum_{n\in\Z}
    \psi_{\Ki{-n }( \BOkk)}( \xp,\xm)
,
\end{align}
where we write $[\BOkk]$ because the function depends on the equivalence
class of $\kp \km$ only. The equivalence relation is given by
\begin{equation}
k=
\begin{pmatrix}
   \kp\\ \km
 \end{pmatrix}
  \equiv
  \Ki{-n } k =
  \begin{pmatrix}
  \kp e^{2\pi\Delta\,n}
    \\ 
  \km e^{-2\pi\Delta\,n}
\end{pmatrix}
.
\label{eq:BO_kpkp_equivalence}
\end{equation}
The previous equation explains the rationale for the parametrization
\eqref{eq:kpkm_parametrization} so that
we can always choose a representative 
\begin{equation}
  0\le \beta < 2 \pi, ~~~ m\ne 0
  ,
\end{equation}
or differently said $\beta\equiv\beta+2\pi$ and therefore we can use the
dual quantum number $l$ using a Fourier transform.
Using the well adapted set of coordinates we can write the spin-0 wave
function in a way to show the natural variables as
\begin{align}
  \Psi_{[\BOkk]}([\xp,\xm])
  &=
\sum_n    
    e^{i\, \tau \left[  \lambda e^{+2\pi\Delta\,n}
    +
    \hspm
    \lambda^{-1}    e^{ -2\pi\Delta\,n} 
    \right]}  
    =
      \hat \Psi(\tau,\lambda, \hspm)
\label{eq:BO_PSI0_tau_lambda}
    .
\end{align}
Again the scalar eigenfunction has a unique equivalence class which mixes
coordinates and momenta.

Now we use the basic trick used in Poisson resummation
\begin{align}
\Psi_{[\BOkk]}([\xp,\xm])
  &=
  \int_{-\infty}^\infty d s~ \delta_P(s)\,
    e^{i\left\{
    \kp  \xp \Lambda^s
    +
    \km  \xm \Lambda^{-s}
    \right\}
    }
    \nonumber\\
  &=
    \frac{1}{2\pi}
    \sum_{l\in\Z}
    \left| \frac{\kp\xp}{\km\xm} \right|^{-i \frac{ l}{2 \loLd} }
    \,
    \int_{-\infty}^\infty d s~ e^{i\, 2\pi l s}
    e^{i\, sgn(\kp\xp) \sqrt{|\kp \km \xp \xm |}
    \left\{
    \Lambda^s
    +
    \nsm \vsm \Lambda^{-s}
    \right\}}
    \nonumber\\
  &=
    \frac{1}{2\pi}
    \sum_{l\in\Z}
    \left( \LLPB{} \right)^{-i \frac{ l}{\loLd} }
    \,
    \int_{-\infty}^\infty d s~ e^{i\, 2\pi l s}
    e^{i\, m t\, 
    \left\{
    \Lambda^s
    +
    \nsm \vsm \Lambda^{-s}
    \right\}}
    \nonumber\\
  &=
    \frac{1}{2\pi}
    \sum_{l\in\Z}
    e^{i l \beta}
    \Biggl[
    e^{i l \vvphi }
    \,
    \int_{-\infty}^\infty d s~ e^{-i\, 2\pi l s}
    e^{i\, m t\, 
    \left\{
    \Lambda^s
    +
    \nsm \vsm \Lambda^{-s}
    \right\}}
    \Biggr]
,    
\end{align}
where the last line represents the change of quantum number from
$m\beta$ to $\ml$ 
and allows us to identify
\begin{equation}
\cNBO
  \tplsi(\tau)=
    \frac{1}{2\pi}
    \,
    \int_{-\infty}^\infty d s~ e^{-i\, 2\pi l s}
    e^{i\, \tau\, 
    \left\{
    \Lambda^s
    +
    \hspm
    \Lambda^{-s}
    \right\}}
,    
\end{equation}
where $\cNBO$ is a constant which depends on the normalization chosen for $\tplsi$. 
This expression gives a integral representation of the o.d.e. solutions.

\subsubsection{Spin 2}
We start with the Minkowskian tensorial wave function where we suppress
all directions but $\xp, \xm$ and $\xx$ since all other directions
behave as $\xx$.
In this case differently from spin 0 we need to keep the dependence on
$\xx$ since it is needed for non trivial physical polarizations since
it enters in the transversality conditions.
Explicitly
\begin{align}
\cNBO
  \psi^{[2]}_{k\,S}(\xp,\xm, \xx)
  &=
    S_{\mu\nu}\, \dd x^\mu  \dd x^\nu\, \psi_k(x)
    \nonumber\\
  &=
    \Biggl[
    S_{++}\, (\dd \xp)^2  
    +
    2
    S_{+-}\, \dd \xp  \dd \xm
    +
    +
    2
    S_{+2}\, \dd \xp  \dd \xx
    \nonumber\\
  &\phantom{=(}
    +
    S_{--}\, (\dd \xm)^2
        +
    2
    S_{-2}\, \dd \xx  \dd \xx
    +
    \nonumber\\
  &\phantom{=(}
    +
    S_{22}\, (\dd \xx)^2
    \Biggr]
    e^{i\left( \kp \xp + \km \xm + \kk \xx \right)}
    ,
\end{align}
which we rewrite in orbifold coordinates 
\begin{align}
\cNBO
  \psi^{[2]}_{k\,S}(t, \vvphi, \xx, \nsm)
  &=
    S_{\alpha\beta}\, \dd x^\alpha  \dd x^\beta\, \psi_k(x)
    \nonumber\\
  \Biggl[
  &{\dd t}^2\,
    \left(2\,S_{+\,-}\,\nsm
    +S_{+\,+}\,e^{2\,\Delta\,\vvphi}
    +S_{-\,-}\,e^ {- 2\,\Delta\,\vvphi }\right)
    \nonumber\\
  +&
     2\,\Delta\,t\,
     \dd t\, \dd\vvphi\,
     \left(S_{+\,+}\,e^{2\,\Delta\,\vvphi}
     -S_{-\,-}\,e^ {- 2\,\Delta\,\vvphi }
     \right)
     \nonumber\\
  +&
     \Delta^2\,t^2
     d\vvphi^2\,
     \left(-2\,S_{+\,-}\,\nsm
     +S_{+\,+}\,e^{2\,\Delta\,\vvphi}
     +S_{-\,-}\,e^ {- 2\,\Delta\,\vvphi }
     \right)
     \nonumber\\
  +&
     2
     \dd t\, {\dd \xx}\,
     \left(S_{-\,2}\,e^ {- \Delta\,\vvphi }\,\nsm
     +S_{+\,2}\,e^{\Delta\,\vvphi}
     \right)
     \nonumber\\
  +& 2 \Delta\,t\,
     \dd \xx\,d\vvphi\,
     \left(S_{+\,2}\,e^{\Delta\,\vvphi}
     -S_{-\,2}\,e^ {- \Delta\,\vvphi }\,\nsm
     \right)
     \nonumber\\
  +&
     ({\dd \xx})^2\,S_{2\,2}
\Biggr]
    e^{i m t \left[ \ \LLPB+
    +
    \hspm \LLPB-
     \right]
     + i \kk \xx}
    .
  \end{align}

  Now we define the tensor wave on the orbifold as a sum over all images as
\begin{align}
\cNBO
  \Psi^{[2]}_{[k\, S]}([x])
  &=
    \sum_n    ( \Ki{ n }d x) \cdot S \cdot ( \Ki{ n }d x)\,
    \psi_{k}( \Ki{ n } x)
    \nonumber\\
  &=
    \sum_n    d x \cdot ( \Ki{ - n }S ) \cdot d x\,
    \psi_{ \Ki{ -n } k}( x)
    .
  \end{align}
  In the last line we have defined the induced action of the Killing vector on
  $(k, S)$ which can be explicitly written as
  \begin{align}
    \Ki{-n}
    \left(
    \begin{array}{c}
      S_{ + + }\\
      S_{ + - }\\
      S_{ - - }\\
      S_{ + 2 }\\
      S_{ -  2}\\
      S_{ 2 2 }\\
    \end{array}
    \right)
    =
    \left(
    \begin{array}{c}
      \LLP{2 n} S_{ + + }\\
      S_{ + - }\\
      \LLP{-2 n} S_{ - - } \\
       \LLP{ n} \Delta S_{ + 2 } \\
      \LLP{-n} S_{- 2} \\
      S_{ 2 2 } 
    \end{array}
    \right)
    ,
    \end{align}
    and it amounts to a trivial scaling.

  In orbifold coordinates
  computing the tensor wave on the orbifold simply
  amounts to sum over all the shifts
  $\vvphi \rightarrow \vvphi + 2\pi n$.
  Then we have to give a close  expression for the sum involving
  powers $ e^{ 2\pi \Delta n}$, explicitly we find  
  \begin{align}
    \sum_n    
    \left( e^{ 2\pi \Delta n}\right)^N
    e^{i\, \tau \left[ \lambda e^{ +2\pi\Delta\,n}
    +
    \hspm\frac{1}{\lambda}
    e^{ -2\pi\Delta\,n} 
    \right]}
    =
    \begin{cases}
      \left[\oh\left( \frac{1}{\lambda} \partial_\tau
        + \frac{1}{\tau}\partial_\lambda  \right)\right]^N
      \hat \Psi(\tau,\lambda, \hspm)
    & N>0
    \\
      \left[\oh\left( {\lambda} \partial_\tau
        - \frac{\lambda^2}{\tau}\partial_\lambda  \right)\right]^N
      \hat \Psi(\tau,\lambda, \hspm)
    & N<0
  \end{cases}
      ,
  \end{align}
  where $\tau$ derivatives higher than $2$ of $\tplsi$ can be reduced with
  the help of the differential equation
  \eqref{eq:BO_eq_diff_tilde_phi}.
  
   We now have to identify the basic polaritazions on the
   orbifold.
   There are three basic observations.
   The quantum number $\beta$ is no longer a good quantum number on
   the orbifold and it is replaced by $l$.
   The relations among orbifold polarizations and Minkowski
   polarizations may depend on $\beta$ as long as the traceless and
   transversality conditions on the orbifold are independent of it.
   These conditions may be a linear combinations of the ones in
   Minkowski.
   Finally it seems reasonable to use the natural variable $\lambda=\LLPB{}$.
   Therefore, as we could guess, we have:
   \begin{align}
&
\cS_{t\,t}=e^ {- 2\,\Delta\,\beta }\,S_{+\,+}
\nonumber\\
&
\cS_{t\,\vvphi}=S_{+\,-}
\nonumber\\
&
\cS_{t\,2}=e^ {- \Delta\,\beta }\,S_{+\,2}
\nonumber\\
&
\cS_{\vvphi\,\vvphi}=e^{2\,\Delta\,\beta}\,S_{-\,-}
\nonumber\\
&
\cS_{\vvphi\,2}=e^{\Delta\,\beta}\,S_{-\,2}
\nonumber\\
&
\cS_{2\,2}=S_{2\,2}
,
\end{align}
which can be trivially inverted as
\begin{align}
&
S_{+\,+}=e^{2\,\Delta\,\beta}\,\cS_{t\,t}
\nonumber\\
&
S_{+\,-}=\cS_{t\,\vvphi}
\nonumber\\
&
S_{+\,2}=e^{\Delta\,\beta}\,\cS_{t\,2}
\nonumber\\
&
S_{-\,-}=e^ {- 2\,\Delta\,\beta }\,\cS_{\vvphi\,\vvphi}
\nonumber\\
&
S_{-\,2}=e^ {- \Delta\,\beta }\,\cS_{\vvphi\,2}
\nonumber\\
&
S_{2\,2}=\cS_{2\,2}
.
\end{align}
When they are inserted into the trace condition they give
\begin{equation}
  \tr(S)=-2 \cS_{t\,\vvphi}+\cS_{2\,2}
,
\end{equation}
while the transversality conditions become
\begin{align}
(k\cdot S)_{+}=
  &
    -e^{\Delta\,\beta}\,\left({m}\,\hspm\,\nsm\,\cS_{t\,t}
    +{m}\,\cS_{t\,\vvphi}-k_{2}\,\cS_{t\,2}\right)
\nonumber\\
(k\cdot S)_{-}=
&
                  -e^ {- \Delta\,\beta }\,\left({m}\,\hspm\,\nsm\,\cS_{t\,\vvphi}
                  +{m}\,\cS_{\vvphi\,\vvphi}-k_{2}\,\cS_{\vvphi\,2}\right)
                 \nonumber\\  
(k\cdot S)_{2}=
  &
-\left({m}\,\hspm\,\nsm\,\cS_{t\,2}+{m}\,\cS_{\vvphi\,2}-k_{2}\,\cS_{2\,2}\right)
,
\end{align}
which are independent of $\beta$ when set to zero.

The final expression for the wave function for the symmetric tensor on
the orbifold reads
\begin{align}
  \Psi^{[2]}_{[k\, S]}([x])
  &=
    \sum_{l\in\Z} e^{i l \beta}
        \Biggl[
    S_{\ml, t t}\, (\dd t)^2  
    +
    2
    S_{\ml, t \vvphi}\, \dd t \dd \vvphi
    +
    +
    2
    S_{\ml,t 2}\, \dd t  \dd \xx
    \nonumber\\
  &\phantom{=(}
    +
    S_{\ml, \vvphi \vvphi}\, (\dd \vvphi)^2
        +
    2
    S_{\ml, \vvphi 2}\, \dd \vvphi  \dd \xx
    +
    \nonumber\\
  &\phantom{=(}
    +
    S_{\ml, 22}\, (\dd \xx)^2
    \Biggr]
,
\end{align}
where the explicit expressions for the components are
\begin{align}
S_{\ml, tt}
=
&+\left[
    -\frac{\tplsi(\tau)\,l\,\lambda^{{{i\,l}\over{\Delta}}}\,\left(l\,\cS_{t\,t}+i\,\Delta\,\cS_{t\,t}+l\,\cS_{\vvphi\,\vvphi}
    -i\,\Delta\,\cS_{\vvphi\,\vvphi}\right)
    }{
    2\,\Delta^2
    }
\right]
{{1}\over{\tau^2}}
\nonumber\\
  &+\left[
    { 1 \over{2\,\Delta} }
    {{{{d}\over{d\,\tau}}\,\tplsi(\tau)\,\lambda^{{{i\,l}\over{\Delta}}}\,
    \left(
    i\,l\,\cS_{t\,t}
    -i\,l\,\cS_{\vvphi\,\vvphi}
    -\Delta\,\cS_{t\,t}
    -\Delta\,\cS_{\vvphi\,\vvphi}
    \right)}}
    \right]
    {{1}\over{\tau}}
\nonumber\\
&+\left[
              \tplsi(\tau)\,\lambda^{{{i\,l}\over{\Delta}}}\,\left(\hspm\,\cS_{t\,t}+2\,\nsm\,\cS_{t\,\vvphi}
              +\hspm\,\cS_{\vvphi\,\vvphi}\right)
\right]
,              
\end{align}
and
\begin{align}
S_{\ml, t \vvphi}
=
&+\left[
    -{
    {\tplsi(\tau)\,l\,\lambda^{{{i\,l}\over{\Delta}}}\,
    \left(l\,\cS_{t\,t}+i\,\Delta\,\cS_{t\,t}-l\,\cS_{\vvphi\,\vvphi}
    +i\,\Delta\,\cS_{\vvphi\,\vvphi}\right)}
    \over
    {2\,\Delta\,{m}}
    }
\right]
{{1}\over{\tau}}
\nonumber\\
&+\left[
              {{{{d}\over{d\,\tau}}\,\tplsi(\tau)\,
              \lambda^{{{i\,l}\over{\Delta}}}\,
              \left(
              i\,l\,\cS_{t\,t}
              -\Delta\,\cS_{t\,t}+
              i\,l\,\cS_{\vvphi\,\vvphi}
              +\Delta\,\cS_{\vvphi\,\vvphi}
              \right)}\over{2\,{m}}}
\right]
\nonumber\\
  &+\left[
    {{\Delta\,
    \hspm\,
    \tplsi(\tau)\,
    \lambda^{{{i\,l}\over{\Delta}}}\,
    \left(\cS_{t\,t}-\cS_{\vvphi\,\vvphi}\right)}\over{{m}}}
    \right]
    \tau
    ,
\end{align}
and
\begin{align}
S_{\ml, \vvphi\vvphi}
=
&+\left[
    -
    {1 \over{2\,{m}^2} }
    {{\tplsi(\tau)\,
    l\,
    \lambda^{{{i\,l}\over{\Delta}}}\,
    \left(
    l\, ( \cS_{t\,t} +\cS_{\vvphi\,\vvphi} )
    +i\,\Delta\, ( \cS_{t\,t} - \cS_{\vvphi\,\vvphi} )
    \right)}
    }
    \right]
\nonumber\\
  &+\left[
    {1 \over{2\,{m}^2} }
    {{\Delta\,
    \left({{d}\over{d\,\tau}}\,\tplsi(\tau)\right)\,
    \lambda^{{{i\,l}\over{\Delta}}}\,
    \left(
    i\,l\,\cS_{t\,t}
    -i\,l\,\cS_{\vvphi\,\vvphi}
    -\Delta\,\cS_{t\,t}
    -\Delta\,\cS_{\vvphi\,\vvphi}
    \right)}}
    \right]
    \tau
\nonumber\\
  &+\left[
    { 1 \over{{m}^2} }
    {{\Delta^2\,
    \tplsi(\tau)\,
    \lambda^{{{i\,l}\over{\Delta}}}\,
    \left(
    \hspm\,\cS_{t\,t}
    +\hspm\,\cS_{\vvphi\,\vvphi}
    -2\,\nsm\,\cS_{t\,\vvphi}
    \right)}}
    \right]
    \tau^2
    ,
\end{align}
and the effectively vector components in the orbifold directions
\begin{align}
S_{\ml, t2}
=
  &+\left[
    {i\over{2\,\Delta}}
    {{\tplsi(\tau)\,
    l\,
    \lambda^{{{i\,l}\over{\Delta}}}\,
    \left(\cS_{t\,2}-\cS_{\vvphi\,2}\,\nsm\right)}}
    \right]
    {{1}\over{\tau}}
\nonumber\\
  &+\left[
    \oh
    {{{{d}\over{d\,\tau}}\,\tplsi(\tau)\,
    \lambda^{{{i\,l}\over{\Delta}}}\,
    \left(\cS_{t\,2}+\cS_{\vvphi\,2}\,\nsm\right)}}
    \right]
    ,
\end{align}
and
\begin{align}
S_{\ml, \vvphi 2}
=
  &+\left[
    {i\over{2\,{m}}}
    {{\tplsi(\tau)\,
    l\,
    \lambda^{{{i\,l}\over{\Delta}}}\,
    \left(\cS_{t\,2}+\cS_{\vvphi\,2}\,\nsm\right)}}
    \right]
\nonumber\\
  &+\left[
    {1\over{2\,{m}}}
    {{\Delta\,
    \left({{d}\over{d\,\tau}}\,\tplsi(\tau)\right)\,
    \lambda^{{{i\,l}\over{\Delta}}}\,
    \left(\cS_{t\,2}-\cS_{\vvphi\,2}\,\nsm\right)}}
    \right]
    \tau
    ,
\end{align}
and finally  the effectively scalar component
\begin{align}
S_{\ml, 22}
=
&
\cS_{2\,2}\,\tplsi(\tau)\,\lambda^{{{i\,l}\over{\Delta}}}
.
\end{align}

\subsection{Overlap of Wave Functions and Their Derivatives
and a Divergent Three Points String Amplitude}
\label{sec:BOoverlap}

Now we consider some overlaps as done for the \NBO\!\!.
The connection between the overlaps on the orbifold and
the sums of images remains unchanged when we change the Killing vector
$\Ki{}$,
hence we can limit ourselves to discuss the integrals on the orbifold space.  

\subsubsection{Overlaps Without Derivatives}
Let us start with the simplest case of the overlap of $N$ scalar wave
functions
\begin{align}
  I^{(N)}
  &=
    \int_{\Omega} \dd^3x\, \sqrt{-\det\, g}
  \prod_{i=1}^N   \Psi_{[\BOkkN{i}]}([\xp,\xm,\xx]))
    \nonumber\\
  &=
    \cNBO^N
     \sum_{ \{\mmN{i}\} \in\Z^N } 
    e^{i \sum_{i=1}^N \mmN{i} \beta_{(i)} }
     \int_{\Omega} \dd^3x\, \sqrt{-\det\, g}
    \prod_{i=1}^N \plsiN{i}
     .
\end{align}
Thsi is always a distribution since the problematic $\mmN{*}=0$ sector
gives a divergence like $(\log(|t|))^N$ around zero.
All other sectors have no issues because of the asymptotics
\eqref{eq:BO_asymtotics}.

\subsubsection{An Overlap With Two Derivatives}
We consider in orbifold coordinates the overlap needed for the amplitude involving two
tachyons and one massive state, i.e.
\begin{align}
  K
  =
  \int_\Omega \dd^3x\,\sqrt{-\det\,g}\,
  &
      g^{\alpha\beta}\,g^{\gamma\delta}\,
      \Psi^{[2]}_{[\kN{3}, \SsN{3}]\, \alpha\gamma}([x])\,
      D_\beta \partial_{\delta} \Psi_{ [\kN{2}] }([x])\,
      \Psi_{ [\kN{1}] }([x])
.
\end{align}
Since we want to use the traceless condition we need to keep
all momenta and polarizations and not only the ones
along the orbifold, then we can write
\begin{align}
  K
  =
  \int_\Omega \dd^3x\,\sqrt{-\det\,g}
  &
    \Bigl[
    +
    \Psi^{[2]}_{[\kN{3}, \SsN{3}]\,  t\, t }\,
    \partial_t^2 \Psi_{ [\kN{2}] }
    \nonumber\\
  &
  -2
  \left(\frac{1}{\Delta t}\right)^2
  \Psi^{[2]}_{[\kN{3}, \SsN{3}]\,  t\, \vvphi }\,
  \left(
  \partial_t \partial_\vvphi \Psi_{ [\kN{2}] }
  -
  \frac{1}{t} \partial_\vvphi \Psi_{ [\kN{2}] }
  \right)
  \nonumber\\
  &
  +
  \left(\frac{1}{\Delta t}\right)^4
  \Psi^{[2]}_{[\kN{3}, \SsN{3}]\,  \vvphi\, \vvphi }\,
  \left(
  \partial_\vvphi^2 \Psi_{ [\kN{2}] }
- \Delta^2 t \partial_t \Psi_{ [\kN{2}] }
  \right)
    \nonumber\\
  &
  -2
  \Psi^{[2]}_{[\kN{3}, \SsN{3}]\,  t\, 2 }\,
  \partial_t \partial_2 \Psi_{ [\kN{2}] }
    \nonumber\\
  &
  +2
  \left(\frac{1}{\Delta t}\right)^2
  \Psi^{[2]}_{[\kN{3}, \SsN{3}]\,  \vvphi\, 2 }\,
  \partial_\vvphi \partial_2 \Psi_{ [\kN{2}] }
    \nonumber\\
  &
  +
  \Psi^{[2]}_{[\kN{3}, \SsN{3}]\,  2\, 2 }\,
  \partial_2^2 \Psi_{ [\kN{2}] }
    \Bigr]
    \Psi_{ [\kN{1}] }
    .
\end{align}

Now consider the behavior for $\mmN{*}=0$ for small $t$.
All the $\partial_\vvphi$ can be dropped since they lower a $\mmN2$.
The leading contributions from spin $2$ components are
$S_{\ml\,t t}\sim \frac{1}{t^2}$,
$S_{\ml\,\vvphi \vvphi}, S_{\ml\,2\,2} \sim 1$
and
$S_{\ml\,t 2}\sim \frac{1}{t}$
therefore the leading $\frac{1}{t^4}$ reads
\begin{align}
  K
  \sim
  \int_{t\sim 0} \dd t\,
  |t|
  &
  \Biggl[
-\oh
  {{d}\over{d\,\tau}}\,\tplsi\,
  ( \cS_{t\,t} + \cS_{\vvphi\,\vvphi} )
  {{1}\over{\tau}}
  \times
  \partial_t^2 \Psi_{ [\kN{2}] }
    \nonumber\\
  &
    +
  \left(\frac{1}{\Delta t}\right)^4
    \times
    {-\Delta^2 \over{2\,{m}^2} }
    {{d}\over{d\,\tau}}\,\tplsi\,
    ( \cS_{t\,t} + \cS_{\vvphi\,\vvphi} )\,
    \tau
    \times
    \left(
    - \Delta^2 t \partial_t \Psi_{ [\kN{2}] }
    \right)    
    \Biggr ]
   \Psi_{ [\kN{3}] }
\end{align}
In the limit of our interest $\Psi_{ [k] }|_{l=0}\sim \tplsi|_{l=0}\sim
log(|t|)$ then the two terms add together because of sign of the
covariant derivative to give
\begin{equation}
  K
  \sim
  \int_{t\sim 0} \dd t\, |t|\,
  \left[
  \left(\oh + \oh \right)
  \frac{\cS_{t\,t} + \cS_{\vvphi\,\vvphi} }{ m^4}
  \frac{\log(|t|)}{ t^4}
  +O\left(   \frac{\log^2(|t|)}{ t}
  \right)
\right]
  ,
  \end{equation}
  which is divergent for the physical polarization
  $\cS_{t\,t}=\cS_{\vvphi\,\vvphi}=- \hspm \nsm \cS_{t\,\vvphi}
  = -\oh \hspm \nsm \cS_{2 2}$.

\begin{center}
{\bf Acknowledgments}    
\end{center}

We thank Marco Bill\'o  and Domenico Orlando for discussions.
  The work is partially supported by the MIUR PRIN Contract 2015 MP2CX4 “Non-perturbative Aspects Of Gauge Theories And Strings”.

\appendix

\section{Complete Tensor Wave Function in the \NBO}
\label{app:NO_tensor_wave}
For the sake of completeness we report the expression of the full \NBO tensor
wave function ($=\frac{l}{\kp}$):
\begin{align}
\begin{pmatrix}
S_{uu}
\\
S_{uv}
\\
S_{uz}
\\
S_{u{i}}
\\
S_{vv}
\\
S_{vz}
\\
S_{v{i}}
\\
S_{zz}
\\
S_{z{i}}
\\
S_{{i}{i}}
\\
\end{pmatrix}
& =~
\cS_{u u}
\begin{pmatrix}
1
\\
0
\\
0
\\
0
\\
0
\\
0
\\
0
\\
0
\\
0
\\
0
\\
\end{pmatrix}\, \pkmkr
+\cS_{u v}
\begin{pmatrix}
{{i}\over{k_{+}\,u}}+{{{L}^2}\over{\Delta^2\,u^2}}
\\
1
\\
{L}
\\
0
\\
0
\\
0
\\
0
\\
0
\\
0
\\
0
\\
\end{pmatrix}\, \pkmkr
+\cS_{u z}
\begin{pmatrix}
{{2\,{L}}\over{\Delta\,u}}
\\
0
\\
\Delta\,u
\\
0
\\
0
\\
0
\\
0
\\
0
\\
0
\\
0
\\
\end{pmatrix}\, \pkmkr
  \nonumber\\
  &
  +\cS_{u {i}}
\begin{pmatrix}
0
\\
0
\\
0
\\
1
\\
0
\\
0
\\
0
\\
0
\\
0
\\
0
\\
\end{pmatrix}\, \pkmkr
+\cS_{v v}
\begin{pmatrix}
-{{3}\over{4\,k_{+}^2\,u^2}}+{{3\,i\,{L}^2}\over{2\,\Delta^2\,k_{+}\,u^3}}+{{{L}^4}\over{4\,\Delta^4\,u^4}}
\\
{{i}\over{2\,k_{+}\,u}}+{{{L}^2}\over{2\,\Delta^2\,u^2}}
\\
{{3\,i\,{L}}\over{2\,k_{+}\,u}}+{{{L}^3}\over{2\,\Delta^2\,u^2}}
\\
0
\\
1
\\
{L}
\\
0
\\
{{i\,\Delta^2\,u}\over{k_{+}}}+{L}^2
\\
0
\\
0
\\
\end{pmatrix}\, \pkmkr
\nonumber\\
  +&\cS_{v z}
\begin{pmatrix}
{{3\,i\,{L}}\over{\Delta\,k_{+}\,u^2}}+{{{L}^3}\over{\Delta^3\,u^3}}
\\
{{{L}}\over{\Delta\,u}}
\\
{{3\,{L}^2}\over{2\,\Delta\,u}}+{{3\,i\,\Delta}\over{2\,k_{+}}}
\\
0
\\
0
\\
\Delta\,u
\\
0
\\
2\,\Delta\,{L}\,u
\\
0
\\
0
\\
\end{pmatrix}\, \pkmkr
+\cS_{v {i}}
\begin{pmatrix}
0
\\
0
\\
0
\\
{{i}\over{2\,k_{+}\,u}}+{{{L}^2}\over{2\,\Delta^2\,u^2}}
\\
0
\\
0
\\
1
\\
0
\\
{L}
\\
0
\\
\end{pmatrix}\, \pkmkr
\nonumber\\
   %
   %
  %
  &+\cS_{z z}
\begin{pmatrix}
{{i}\over{k_{+}\,u}}+{{{L}^2}\over{\Delta^2\,u^2}}
\\
0
\\
{L}
\\
0
\\
0
\\
0
\\
0
\\
\Delta^2\,u^2
\\
0
\\
0
\\
\end{pmatrix}\, \pkmkr
+\cS_{z {i}}
\begin{pmatrix}
0
\\
0
\\
0
\\
{{{L}}\over{\Delta\,u}}
\\
0
\\
0
\\
0
\\
0
\\
\Delta\,u
\\
0
\\
\end{pmatrix}\, \pkmkr
+\cS_{{i} {j}}
\begin{pmatrix}
0
\\
0
\\
0
\\
0
\\
0
\\
0
\\
0
\\
0
\\
0
\\
\delta_{i j}
\\
\end{pmatrix}\, \pkmkr
.
\end{align}

\section{Complete Overlap With Two Derivatives in the \NBO}
\label{app:NO_full_TTS}
For the sake of completeness we report the full expression of the overlap
with two derivatives considered in the main text which corresponds to
the colour ordered amplitude of two tachyons and one level 2 massive
state:
\begin{equation}
    \begin{split}
        K = 
        \cN^2
        \int \dd^D x\, \sqrt{-\det g}\,
            &
            \Bigg[
                \mathfrak{s}^{(-3)}
                \left(
                    \kmkrN{i}_{i = 1,\,2,\,3},~
                    \{ \mathcal{S} \}
                \right)
                u^{-3}
                \\
                &
                +
                \mathfrak{s}^{(-2)}
                \left(
                    \kmkrN{i}_{i = 1,\,2,\,3},~
                    \{ \mathcal{S} \}
                \right)
                u^{-2}
                \\
                &
                +
                \mathfrak{s}^{(-1)}
                \left(
                    \kmkrN{i}_{i = 1,\,2,\,3},~
                    \{ \mathcal{S} \}
                \right)
                u^{-1}
                \\
                &
                +
                \mathfrak{s}^{(0)}
                \left(
                    \kmkrN{i}_{i = 1,\,2,\,3},~
                    \{ \mathcal{S} \}
                \right)
                \\
                &
                +
                \mathfrak{s}^{(1)}
                \left(
                    \kmkrN{i}_{i = 1,\,2,\,3},~
                    \{ \mathcal{S} \}
                \right)
                u
            \Bigg]~
            \prod_{i = 1}^3 \pkmkrN{i}
    \end{split}
\end{equation}
where:
\begin{eqnarray}
    &
    \mathfrak{s}^{(-3)}
    & =
    \Bigg(
        -{{\KKK{2}{+}^4\,{\mmN{3}}^4-4\,\KKK{2}{+}^3\,\KKK{3}{+}\,
        {\mmN{2}}\,{ \mmN{3}}^3}\over{4\,\KKK{2}{+}^2\,\KKK{3}{+}^4\,\Delta^3}}
        \nonumber\\
        &&
        -{{6\,\KKK{2}{+}^2\,\KKK{3}{+}^2\,{
        \mmN{2}}^2\,{ \mmN{3}}^2+\KKK{3}{+}^4\,{\mmN{2}}^4}\over{4\,\KKK{2}{+}^2\,\KKK{3}{+}^4\,\Delta^3}}
    \Bigg)\,
    \cS_{v\,v},
\end{eqnarray}
\begin{eqnarray}
    &
    \mathfrak{s}^{(-2)}
    & =
        \Biggl(
     -{
    {i\,\left(
    3\,\KKK{2}{+}^2\,\KKK{3}{+}\,{\mmN{3}}^2
    +3\,\KKK{2}{+}^3\,{\mmN{3}}^2
    -4\,\KKK{2}{+}\,\KKK{3}{+}^2\,{ \mmN{2}}\,{\mmN{3}}
    -6\,\KKK{2}{+}^2\,\KKK{3}{+}\,{ \mmN{2}}\,{\mmN{3}}
    \right)
     }
     \over{2\,\KKK{2}{+}\,\KKK{3}{+}^3\,\Delta}
     }
     \nonumber\\
&& +
  \frac{
  -i\,\left(
  +3\,\KKK{3}{+}^3\,{\mmN{2}}^2
  +3\,\KKK{2}{+}\,\KKK{3}{+}^2\,{\mmN{2}}^2\right)
  }{
  2\,\KKK{2}{+}\,\KKK{3}{+}^3\,\Delta
  }
        \Biggr) \,
        \cS_{v\,v} 
\nonumber\\
&& +\left( -{{{ \mmN{3}}\,\left(\KKK{2}{+}^2\,{
              \mmN{3}}^2-3\,\KKK{2}{+}\,\KKK{3}{+}\,{ \mmN{2}}\,{
              \mmN{3}}+3\,\KKK{3}{+}^2\,{
              \mmN{2}}^2\right)}\over{\KKK{3}{+}^3\,\Delta^2}}
              \right)
              \,
              \cS_{v\,z},
\end{eqnarray}
\begin{eqnarray}
    &
    \mathfrak{s}^{(-1)}
    & =
        \left( -{{\left(\KKK{2}{+}\,{ \mmN{3}}-\KKK{3}{+}\,{
              \mmN{2}}\right)^2}
              \over{\KKK{3}{+}^2\,\Delta}} \right)
              \,
              \cS_{u\,v} 
         \nonumber\\
&& +\Biggl( -{{2\,\KKK{2}{+}^2\,{ \mmN{3}}^2\,(r_{(2)}+\vk_{(2)}^2
              )\,
              +2\,\KKK{3}{+}^2\,{ \mmN{2}}^2\,(r_{(2)}+\vk_{(2)}^2
              )\,
              -8\,\KKK{2}{+}^3\,\KKK{3}{+}\,{ \mmN{2}}\,{
              \mmN{3}}
         }
              \over{4\,\KKK{2}{+}^2\,\KKK{3}{+}^2\,\Delta}}
\nonumber\\
  && - \frac{
              -3\,\KKK{2}{+}^2\,\KKK{3}{+}^2\,\Delta^2
              -6\,\KKK{2}{+}^3\,\KKK{3}{+}\,\Delta^2
              -3\,\KKK{2}{+}^4\,\Delta^2
    }{
    4\,\KKK{2}{+}^2\,\KKK{3}{+}^2\,\Delta
    }
              \Biggr) \,
              \cS_{v\,v} 
\nonumber\\
&& +\left( -{{i\,\left(3\,\KKK{2}{+}\,\KKK{3}{+}\,{
              \mmN{3}}+3\,\KKK{2}{+}^2\,{ \mmN{3}}-2\,\KKK{3}{+}^2\,{
              \mmN{2}}-3\,\KKK{2}{+}\,\KKK{3}{+}\,{
              \mmN{2}}\right)}\over{\KKK{3}{+}^2}} \right) \,
              \cS_{v\,z} 
\nonumber\\
  && +\left( {{\KKK{2}{i}\,{
    \mmN{3}}\,\left(\KKK{2}{+}\,{
              \mmN{3}}
              -2\,\KKK{3}{+}\,{
              \mmN{2}}\right)}
              \over{
              \KKK{3}{+}^2\,\Delta}
              } \right) \,
              \cS_{v\,{i}} 
\nonumber\\
&& +\left( -{{\left(\KKK{2}{+}\,{ \mmN{3}}-\KKK{3}{+}\,{
              \mmN{2}}\right)^2}\over{\KKK{3}{+}^2\,\Delta}} \right)
              \,
              \cS_{z\,z},
\end{eqnarray}
\begin{eqnarray}
    &
    \mathfrak{s}^{(0)}
    & =
    \left(
             -{{i\,\KKK{2}{+}\,\left(\KKK{3}{+}+\KKK{2}{+}\right)\,\Delta}\over{\KKK{3}{+}}} \right) \,
             \cS_{u\,v} 
\nonumber\\
&& +\left( -{{2\,\KKK{2}{+}\,\left(\KKK{2}{+}\,{ \mmN{3}}-\KKK{3}{+}\,{
              \mmN{2}}\right)}\over{\KKK{3}{+}}} \right) \,
              \cS_{u\,z} 
\nonumber\\
&& +\left(
              -{{i\,\left(\KKK{3}{+}+\KKK{2}{+}\right)\,\Delta\,(r_{(2)}+\vk_{(2)}^2 )\,}\over{2\,\KKK{2}{+}\,\KKK{3}{+}}} \right) \,
              \cS_{v\,v} 
\nonumber\\
&& +\left( -{{{ \mmN{3}}\,(r_{(2)}+\vk_{(2)}^2
              )\,-2\,\KKK{2}{+}\,\KKK{3}{+}\,{
              \mmN{2}}}\over{\KKK{3}{+}}} \right) \,
              \cS_{v\,z} 
              \nonumber\\
&& +\left( {{i\,\KKK{2}{i}\,\KKK{2}{+}\,\Delta}\over{\KKK{3}{+}}}
              \right) \,
              \cS_{v\,{i}} 
\nonumber\\
&& +\left(
              -{{i\,\KKK{2}{+}\,\left(\KKK{3}{+}+\KKK{2}{+}\right)\,\Delta}\over{\KKK{3}{+}}} \right) \,
              \cS_{z\,z} 
              \nonumber\\
&& +\left( {{2\,\KKK{2}{i}\,\left(\KKK{2}{+}\,{ \mmN{3}}-\KKK{3}{+}\,{
          \mmN{2}}\right)}\over{\KKK{3}{+}}} \right) \,\cS_{z\,{i}},
\end{eqnarray}
\begin{eqnarray}
    &
    \mathfrak{s}^{(1)}
    & =
\left( -\KKK{2}{+}^2\,\Delta \right) \,\cS_{u\,u} 
\nonumber\\
&& +\left( -\Delta\,(r_{(2)}+\vk_{(2)}^2 )\, \right) \,\cS_{u\,v} 
\nonumber\\
  && +\left( 2\,\KKK{2}{i}\,\KKK{2}{+}\,\Delta \right) \,\cS_{u\,{i}} 
\nonumber\\
&& +\left( -{{\Delta\,(r_{(2)}+\vk_{(2)}^2 )\,^2}\over{4\,\KKK{2}{+}^2}} \right) \,\cS_{v\,v} 
              \nonumber\\
&& +\left( 2\,\KKK{2}{i}\,\KKK{2}{+}\,\Delta \right) \,\cS_{v\,{i}} 
          \nonumber\\
  && +\left(
    -
    \KKK{2}{i}\KKK{2}{j}\,\Delta \right) \,\cS_{{i}\,{j}}.
\end{eqnarray}

\bibliographystyle{unsrt}
\bibliography{time_dependent_bck}

\end{document}